\documentclass[12pt]{article}
\usepackage{putex}
\usepackage{graphicx}
\usepackage{color}
\usepackage{fontenc}
\usepackage{epsfig}

\def\B{{B}}
\def\Ff{{ F}}

\usepackage{amsmath, amsthm}
\def\ben{\begin{enumerate}} \def\een{\end{enumerate}}
\def\beq{\begin{equation}} \def\eeq{\end{equation}}
\def\beqn{\begin{equation*}} \def\eeqn{\end{equation*}}
\def\bea{\begin{eqnarray}} \def\eea{\end{eqnarray}}
\def\ba{\begin{array}} \def\ea{\end{array}}
\def\beann{\begin{eqnarray*}} \def\eeann{\end{eqnarray*}}
\def\beasn{\begin{sneqnarray}} \def\eeasn{\end{sneqnarray}}

\def\nn{\nonumber}
\def\pa{\partial}
\def\m{\mu}
\def\be{\begin{equation}} \def\ee{\end{equation}}

   \def\g{\gamma} \def\G{\Gamma}
\def\d{\delta} \def\D{\Delta} 
   
 \def\l{\lambda} \def\L{\Lambda} \def\m{\mu} \def\n{\nu}
  \def\p{\pi}  
\def\s{\sigma}

\begin{document}
\thispagestyle{empty}
\begin{flushright}\scshape

UB-ECM-PF-11/57

ICCUB-11-157
\end{flushright}
\vskip1cm

\begin{center}

{\LARGE\scshape
Chiral effective theory with a light scalar and lattice QCD
\par}

\vskip15mm

\textsc{ J. Soto$^{ab}$, P. Talavera$^{bc}$ and J. Tarr\'us$^{ab}$}
\par\bigskip
$^a${\em
Departament d'Estructura i Constituents de la Mat\`eria,
Universitat de Barcelona,\\
Diagonal 647, E-08028 Barcelona, Catalonia, Spain.}\\[.1cm]
$^b${\em
Institut de Ci\`encies del Cosmos,
Universitat de Barcelona,\\
Diagonal 647, E-08028 Barcelona, Catalonia, Spain.}\\[.1cm]
$^c${\em
Departament de F{\'\i}sica i Enginyeria Nuclear,
Universitat Polit\`ecnica de Catalunya,\\
Comte Urgell 187, E-08036 Barcelona, Spain.}\\[.1cm]
\vspace{5mm}
\end{center}

\section*{Abstract}

We extend the usual chiral perturbation theory framework ($\chi$PT) to allow the inclusion of a light dynamical isosinglet scalar. Using lattice QCD results, and a few phenomenological inputs, we explore the parameter space of the effective theory. We discuss the S--wave pion--pion scattering lengths, extract the average value of the two light quark masses and evaluate the impact of the dynamical singlet field in the low--energy constants $\bar{l}_1$, $\bar{l}_3$ and $\bar{l}_4$ of $\chi$PT. We also show how to extract the mass and width of the sigma resonance from chiral extrapolations of lattice QCD data. 

\bigskip

PACS: 12.39.Fe, 14.40.Be
\vspace{3mm} \vfill{ \hrule width 5.cm \vskip 2.mm {\small
\noindent E-mails: joan.soto@ub.edu, pere.talavera@icc.ub.edu, tarrus@ecm.ub.es }}

\newpage
\setcounter{page}{1}

\section{Introduction}

Chiral Perturbation Theory, $\chi$PT \cite{Weinberg:1978kz,Gasser:1983yg}, has become a standard tool for the phenomenological description of QCD processes involving pseudo--Goldstone bosons at low--energy (see \cite{Ecker:1995zu} for a review). It is grounded in a few simple assumptions: (i) the underlying theory of strong interactions, namely QCD, has an exact $SU(n_f)\times SU(n_f)$ chiral symmetry in the limit of vanishing $n_f$ light quark masses that is spontaneously broken down to $SU(n_f)$, (ii) there is a mass gap ($\sim \Lambda_\chi$) for all states except for the Goldstone bosons, and (iii) the exact chiral symmetry is explicitly broken by the actual non--vanishing quark masses, $m_q\ll \Lambda_\chi$. Under those assumptions one can construct a low--energy effective theory, $\chi$PT, for the $n_f$ pseudo--Goldstone bosons organized in powers of $p / \Lambda_\chi$ and $m_q / \Lambda_\chi$, where $p$ is the typical momentum of the low--energy  processes ($p \ll \Lambda_\chi$). In practice, $ \Lambda_\chi$ is taken of the order of the rho mass ($m_\rho\sim 770\, {\rm MeV}$), and the pseudo--Goldstone bosons are identified with the pions for $n_f=2$ ($m_\pi\sim 140\, {\rm MeV}$) and with the lightest octet of pseudoscalar mesons for $n_f=3$, which also includes the kaons ($m_K\sim 490\, {\rm MeV}$) and the eta ($m_\eta\sim 550\, {\rm MeV}$).

Scattering amplitudes can be systematically calculated within this framework to a given order in $p^2\sim m_\pi^2$ over $\Lambda_\chi^2$ . However, when pion scattering amplitudes are calculated in the isoscalar channel, a bad convergence is observed, even at reasonably low--momenta. This has led some authors to resum certain classes of diagrams, using a number of unitarization techniques (see, for instance, \cite{Dobado:1989qm,Dobado:1996ps,Oller:1997ti,Oller:1997ng,Oller:1998hw}). Most of these approaches improve considerably the description of data with respect to standard $\chi$PT, and indicate that a scalar isospin zero resonance at relatively low--mass, the sigma, exist. In fact the mass and width of the sigma resonance are nowadays claimed to be known very accurately $m_\sigma= 441^{+16}_{-8}\,{\rm MeV}\,, \Gamma /2= 272^{+9}_{-12.5}\,{\rm MeV}$ \cite{Caprini:2005zr,Leutwyler:2008xd} (see also \cite{GarciaMartin:2011jx}).

Under the $SU(3)$ perspective one may find surprising that the effective theory contains kaons but not other states with similar masses, but different quantum numbers, that can be equally excited in a collision at intermediate stages. The relatively low--mass of the sigma resonance, with respect to the chiral cutoff, $\Lambda_\chi$, and its proximity to the value of the kaon mass suggests that it may be convenient to introduce it as an explicit degree of freedom in an extension of $\chi$PT, thus lifting the assumption (ii) above. It is in fact an old observation by Weinberg \cite{Weinberg:1963zz}, that the explicit inclusion of resonances in a Lagrangian generically improves perturbation theory.  

We implement this observation here in a chiral effective theory framework that involves a dynamical singlet field together with the lowest pseudo--Goldstone bosons. We write down the most general $SU(2)$ chiral Lagrangian including an isospin zero scalar field at order $p^4$ and calculate a number of observables at this order. We show that for a large scalar mass, the effect of the scalar  reduces to just redefinitions of the low--energy constants (LEC), hence explicitly demonstrating that our approach is compatible with standard  $\chi$PT. However if we count the mass of the scalar as 
order $p^2$, namely of the same size as the pion mass, the non--analytic pieces of our amplitudes differ from those 
of $\chi$PT. Furthermore, the quark mass dependence of the observables is also different. We compare this effective 
theory, which we call $\chi$PT$_S$, versus standard $\chi$PT against lattice data on the pion mass $m_\pi$ and the 
pion decay constant $F_\pi$ 
\cite{Baron:2009wt}, and on the pion--pion S-wave scattering lengths \cite{Fu:2011bz}. At the current precision 
the lattice data is unable to tell apart $\chi$PT 
from $\chi$PT$_S$. 

We organize the paper as follows. In the next section we discuss the power counting, construct the Lagrangian up to next--to--leading order (NLO), and compare it with the one of the linear sigma model. In section \ref{ax2pf} and section \ref{s2p} we calculate the two--point function of the axial--vector current and of the scalar field respectively, up to NLO.  In section \ref{matchlatdat} we perform a number of fits to lattice data for $m_\pi$ and $F_\pi$ both in $\chi$PT and in $\chi$PT$_S$ in order to constrain the parameter space of the latter. In section \ref{swpipisl} we discuss the LO S--wave scattering lengths, and compare the results of $\chi$PT$_S$ together with those of $\chi$PT with very recent lattice data. In particular it is shown how the mass and decay width of the sigma resonance can be extracted from them. We close with a discussion of our results and the conclusions in section \ref{discon}.

\section{Lagrangian and power counting}

Our aim is to construct an effective field theory containing pions and a singlet scalar field as a degrees of freedom, that holds for processes involving only low--energy pions as the asymptotic states 
\be
\label{range}
p \,,\, m_\pi (\sim 140\, {\rm MeV}) \,,\, {m_{S}} (\sim 440\, {\rm MeV}) \ll \Lambda_{\chi}. 
\ee
The structure of the effective Lagrangian will be independent of the underlying mechanism of spontaneous chiral symmetry 
breaking. It consists of an infinite tower of chiral invariant monomials combining pions and a singlet scalar field with the generic appearance
\be
\label{eee}
{\cal L}^{\rm eff} = \sum_{(k,l,r)} {\cal L}_{(k,l,r)}\,,
\ee
where ${\cal L}_{(k,l,r)}$ contains $k$ powers of derivatives, $l$ powers of the scalar or pseudoscalar sources and 
finally $r$ powers of the singlet field. 
\be
\label{counttwo}
{\cal L}_{(k,l,r)}\sim \Lambda_\chi^4\left(\frac{p}{\Lambda_\chi}\right)^{k+r} \left(\frac{m_{\rm q}}{\Lambda_{\chi}}\right)^l \,,
\ee
being $p$ a typical momentum in the process that we have assumed to be of the order of the scalar particle mass. One possible manner to relate these scales is to assume that $p^2\sim m_\pi^2\sim m_{\rm q}\Lambda_\chi$, like in standard $\chi$PT. Hence, in the chiral counting ${\cal L}_{(k,l,r)}$ is of order $p^{k+r+2l}$. In fact, in this paper we only use the inequalities in (\ref{range}). More refined hierarchies, like $m_\pi \ll {m_{S}} \,,\, p \ll \Lambda_{\chi}$ may be interesting to explore in the future. Notice that terms with $k+r+l < 4$ correspond to relevant operators and, hence, their dimensionful constant may be tuned to a scale smaller than the natural one $\Lambda_\chi$, as it happens in standard $\chi$PT ($F_\pi \ll \Lambda_\chi$).

The Lagrangian involving pions and scalar fields transforming as a singlet under $SU(2)_R\times SU(2)_L$, respecting Chiral symmetry, $P$ and $C$ invariance, has been presented in the linear approximation in \cite{Ecker:1988te,Cirigliano:2006hb} and up to quadratic terms in \cite{Rosell:2005ai}. For the time being, we will collect only the relevant terms necessary for our purposes. The leading order (LO) Lagrangian consist of three parts: the standard Goldstone boson Chiral Lagrangian, that we do not discuss further, terms involving the scalar field only, and interaction terms between the scalar field and the pseudo-Goldstone bosons.

\subsection{Leading Lagrangian}

Consider first the part of  ${\cal L}^{\rm eff}$ containing only the singlet scalar field. In the absence of any symmetry hint we are forced to write the most general polynomial functional,
\bea
{\cal L}^{S}
&=&\frac{1}{2}\pa_{\mu}{S} \pa^{\mu}{S}-{1\over 2}{\mathaccent 23 m}^2_{S}{S}{S}-\l_1 S-\frac{\l_3}{3!}{S}^3-\frac{\l_4}{4!}{S}^4+\cdots
\label{ss}
\eea
where the dots indicate terms suppressed by powers of $1/\Lambda_{\chi}$. Suppose that we deal with the chiral limit. At LO $\l_1$ must be set to zero in order to avoid mixing of $S$ with the vacuum, and at higher orders it must be adjusted for the same purpose. The mass and the coupling constants above are functions of the small scale ${\mathaccent 23 m}_{S}$ and the large scale $\Lambda_{\chi}$, ($ {\mathaccent 23 m}_{S}\ll \Lambda_{\chi}$). 
Their natural values would be $\l_3\sim {\cal O}(\Lambda_{\chi})$ and $\l_4 \sim {\cal O}(1)$. In that case, the scalar sector above becomes strongly coupled. However, strongly coupled scalar theories in four dimensions are believed to be trivial \cite{Luscher:1987ay,Frohlich:1982tw}. Their exact correlation functions factorize according to Wick's theorem and consequently they behave as if the theory were non--interacting. A practical way of taking this fact into account is just setting $\l_3=\l_4=0$, which we will do in the following. When the interactions of the scalar with the pseudo--Goldstone bosons are taken into account, small (${\mathaccent 23 m}_S^2/\Lambda_{\chi}^2$ suppressed) but non--vanishing values of $\l_3$ and $\l_4$ are required to ensure perturbative renormalization of the whole ${\cal L}^{\rm eff}$. 

The second contribution we are interested in is the lowest order Lagrangian describing the interaction of the scalar field with the pseudo--Goldstone bosons. As a basic building block we use the unitary matrix $U(x)$ to parameterize the Goldstone boson fields, that may be taken as,
\be
U = e^{i\phi /F}=u^2\,, \quad 
\phi=\left( \begin{array}{cc}
\pi^0 & \sqrt{2} \pi^+\\
\sqrt{2} \pi^- & -\pi^0
\end{array}
\right)\,,
\label{pions}
\ee
although final results for observable quantities do not depend on this specific choice. At LO, $F$ may be identified 
with the pion decay constant $F_\pi$. We also use the building block,
\be
\label{blocks}
\chi  = s+ip= 2B \hat{m}\Identity \,.
\ee
On the r.h.s. we have set the pseudoscalar source, $p$, equal to zero and $s$ to the diagonal matrix. As we will work in the isospin limit we will use $\hat{m}$ referring to the average quark mass between $m_u$ and $m_d$. The covariant derivative acting on $U(x)$ is defined as usual, containing the external vector and axial sources, $D_\mu U =  \partial_\mu U - i [v_\mu,U] + i \left\{ a_\mu,U\right\} \,.$ The transformation laws for all these building blocks under the local symmetry group $G=SU(2)_R\times SU(2)_L$ are dictated by
\eqn{sylaw}{
U\,{\stackrel{G}{\to}}\, g_R\, U\, g_L^{-1}\,,\quad D_\mu U\,{\stackrel{G}{\to}}\, g_R\, D_\mu U\, g_L^{-1}\,,\quad
S\, {\stackrel{G}{\to}}\,  S\,,\quad
s+ip\, {\stackrel{G}{\to}}\, g_R\,(s+ip)\, g_L^{-1}\,,\cr
r_\mu=v_\mu+a_\mu\,{\stackrel{G}{\to}}\, g_R\,r_\mu\, g_R^{-1}+ i g_R \partial_\mu g_R^{-1}  \,,\quad
l_\mu=v_\mu-a_\mu\,{\stackrel{G}{\to}} \,g_L\,l_\mu\, g_L^{-1} + i g_L \partial_\mu g_L^{-1} \,,
}
where $g_R, g_L \subset SU(2)$\,. 

With all those ingredients one can construct the Lagrangian,
\bea
\label{sigmapion}
{\cal L}^{(2)} = &&
\left( \frac{\Ff^2}{4} +\Ff c_{1d} S+ c_{2d} S^2+\cdots \right) \langle D_\mu U D^\mu U^\dagger\rangle\cr &&+
\left( \frac{\Ff^2}{4} +\Ff c_{1m} S+ c_{2m} S^2+\cdots\right)  \langle \chi^\dagger U + \chi U^\dagger\rangle
\eea
where the ellipsis stand for higher order terms involving more powers of the singlet field (or derivatives on them), which are suppressed by powers of $1/\Lambda_{\chi}$. 

At this point a small digression is in order; notice the peculiarity of (\ref{sigmapion}) with respect to the usual chiral expansion. At this order, both expansions can be cast in the form,
\eqn{expone}{
{\cal L} \sim \sum_{k} b_{k}(\Lambda_{\chi},S) {\mathbb O}^{(k)} 
}
${\mathbb O}^{(k)}$ being an operator of order $k$ including only the pseudo--Goldstone bosons and $b_k(\Lambda_{\chi},S)$ its corresponding ``Wilson coefficient'', that can depend on the singlet field if one considers the theory with the scalar field inclusion. While in the standard theory the power counting is given entirely by the operator, i.e. $b_k(\Lambda_{\chi})\sim  {\cal O}(\Lambda_{\chi}^{4-k})$, in the  extended version one also has to take into account that the Wilson coefficients themselves have a power expansion in $S/\Lambda_{\chi}$. At higher orders operators containing the derivatives of the scalar field must also be included.
 
Before closing this section we would like to remark that even if we have kept for $F$ and $B$ the same names as in $\chi$PT, they are now parameters of a different theory and, hence, their values are expected to differ from those in $\chi$PT.

\subsection{Comparison with the linear sigma model}

Hitherto we have included in a dynamical fashion a scalar particle interacting with pseudo--Goldstone bosons. One may wonder if there is any relation between the effective theory just introduced and the old linear sigma model \cite{GellMann:1960np} (see \cite{Gasiorowicz:1969kn} for a review), which we discuss next. The starting point for the construction of the linear-$\sigma$ model is an $O(4)$ invariant action. The global $O(4)$ symmetry is spontaneously broken down to $O(3)$ because the scalar field develops a non--zero vacuum expectation value $v$. 

The Lagrangian reads 
\eqn{sig}{
{\cal L}_\sigma = \frac{1}{2} \left(\partial_\mu \sigma \partial^\mu\sigma+\partial_\mu\vec{\varphi} \partial^\mu\vec{\varphi}\right) -\frac{\lambda_\s}{4} \left(\sigma^2 + \vec{\varphi}^2 - v^2\right)^2\,,
}
where $\vec{\varphi}$ is an isotriplet pseudoscalar field, usually identified with the pion, and $\sigma$ is an isosinglet scalar field that, after the shift $\sigma \to \sigma + v$ is usually identified with the sigma resonance. We do not display the part of the model containing nucleons because 
it has no relevance for our discussion. Since $O(4) \cong SU(2)\times SU(2)$ for group elements close to the identity, the model transforms correctly under the $SU(2)_R\times SU(2)_L$ chiral symmetry of two--flavor QCD with massless quarks. To see this explicitly we make the change $\Sigma = \sigma \Identity - i \vec{\tau}\cdot \vec{\varphi}$, being $\vec{\tau}$ the Pauli matrices. Then (\ref{sig}) can be written as
\eqn{sig2}{
{\cal L}_\sigma = \frac{1}{4} \langle \partial_\mu \Sigma \partial^\mu\Sigma^\dagger\rangle - \frac{\lambda_\s}{16} \left(\langle \Sigma^\dagger \Sigma\rangle-2 v^2\right)^2\,,
}
that explicitly exhibits the desired symmetry (\ref{sylaw}), if we transform $\Sigma\,{\stackrel{G}{\to}}\, g_R\, \Sigma\, g_L^{-1}$ \cite{Ecker:1995zu}.
The traditional identification of the $\vec \varphi$ fields with the pions and the $\sigma$ field (after the shift) with the sigma resonance, which is fine concerning the transformations under the unbroken subgroup $O(3)\cong SU(2)$, becomes problematic if one wishes to implement the non--linear $SU(2)_R\times SU(2)_L$ symmetry that the model retains after the shift $\sigma \to \sigma + v$ is performed. In order to make the non--linear $SU(2)_R\times SU(2)_L$ symmetry manifest in the Lagrangian above and keep the transformations of the Goldstone bosons in the standard way \cite{Coleman:1969sm,Callan:1969sn}, as we have done in the previous section, it is convenient to perform a polar decomposition of $\Sigma$, $\Sigma = (v+S) U\,,$ with $U$ being a unitary matrix collecting the phases, to be identified with the $U$ appearing in (\ref{pions}), and $S$ a real scalar field, to be identified with our singlet field above. We remark that $S$ must not be mistaken by the $\sigma$ field in the original variables of the linear sigma model. The symmetry transformations of the fields $S$ and $U$ are the same as in (\ref{sylaw}). This change of variables leads to 
\eqn{sig3}{
{\cal L}_\sigma = \left(\frac{v^2}{4} +\frac{v}{2} S+ \frac{1}{4} S^2 \right) \langle D_\mu U D^\mu U^\dagger\rangle
+\frac{1}{2}\partial_\mu S\partial^\mu S-\lambda_\s v^2\left( S^2 + \frac{S^3}{v} +\frac{S^4}{4 v^2}\right)\,.
}
The terms with covariant derivatives above have the very same functional form as the terms with derivatives of (\ref{sigmapion}), with the identifications $v=F$, $c_{1d}=1/2$ and $c_{2d}=1/4$. However, the terms with no derivatives, the potential, are set to zero (or, at higher orders, to small values uncorrelated to the rest of the parameters) in $\chi$PT$_S$, except for the mass term, for which ${\mathaccent 23 m}_{S}^2=2\lambda_\s v^2$. 
This is because the underlying mechanism of chiral symmetry breaking is assumed to take place at the scale $\Lambda_\chi$,  and hence it must not be described in the effective theory.

Since pions are not massless in nature, a small explicit breaking of the $O(4)$ symmetry had to be introduced. This was traditionally done by adding a term $\delta {\cal L}_\sigma = H \sigma$. In terms of the new variables this term reads
\be
\delta {\cal L}_\sigma = H \sigma = \frac{H}{4}\langle \Sigma +\Sigma^\dagger\rangle = \frac{H}{4}(v+S)\langle U +U^\dagger\rangle\, .
\ee
Hence, it has exactly the same functional form as the terms with no derivatives in (\ref{sigmapion}), once $\chi$ is set to $2B\hat m\Identity$, with the identifications $H=2 F B\hat m$, $c_{1m}=1/4$ and $c_{2m}=0$.

In summary, the Lagrangian of $\chi$PT$_S$ at LO  differs from the one of the linear sigma model {\sl only} in two respects: (i) the self--interactions of the scalar field $S$ are set to zero (or, at higher orders, to small values uncorrelated to the rest of the parameters), and (ii) it has four additional free parameters controlling the interaction of the scalar field $S$ with the pions: $c_{1d},\,c_{2d},\,c_{1m},\,c_{2m}$.

\subsection{Chiral symmetry constraints}\label{chisymcon}

To envisage the effects of explicit chiral symmetry breaking on the dynamics of the singlet field we set $U$ to the vacuum configuration  ($U=\Identity$). The terms proportional to the quark masses in (\ref{sigmapion}) induce new terms in the Lagrangian of $S$, that can be reshuffled into the coefficients of (\ref{ss}). For the first two terms one finds explicitly
\be
\l_1 \to \lambda_1 - 8 \Ff c_{1m} \B \hat{m} \,,\quad  {\mathaccent 23 m}_{S}^2 \to m_{S}^2 = {\mathaccent 23 m}_{S}^2 - 16  c_{2m} \B \hat{m}\,.
\label{spredef}
\ee 
As a consequence the singlet field is brought out of its minimum in the chiral limit by terms proportional to $\hat{m}$. Hence, the direct consequence of the inclusion of non--vanishing quark masses results in a new contribution to the singlet--vacuum mixing. The new scalar field describing the first excitation with respect to the vacuum may be obtained by carrying out the following shift
\be
\label{shif}
S\rightarrow S+\Ff S_0\, \quad {\rm with} \quad  S_0=8 c_{1m} \frac{ \B \hat{m}}{m_{S}^2}-\frac{\lambda_1}{m^2_S\Ff}\,.
\ee
After this shift, and upon separating the vacuum contribution, the original Lagrangian (\ref{sigmapion}) keeps essentially the same form,
\bea
\label{newsigmapion}
{\cal L}^{(2)} = 
&&\left( \frac{\Ff^2}{4} r_{0d}+\Ff r_{1d} S+ r_{2d} S^2+\cdots \right) \langle D_\mu U D^\mu U^\dagger\rangle+\nonumber \\
&&\left( \frac{\Ff^2}{4} r_{0m}+\Ff r_{1m} S+ r_{2m} S^2+\cdots\right) \left( \langle \chi^\dagger U + \chi U^\dagger\rangle -\langle \chi^\dagger + \chi \rangle\right)\,, 
\eea
provided we redefine the LEC as
\eqn{redef}{
\begin{tabular}{ll}
 $r_{0d}=  1+ 4c_{1d} S_0 +4 c_{2d} S_0^2+\ldots\,,$ &  $r_{0m}=  1+ 4c_{1m} S_0 +4 c_{2m} S_0^2+\ldots\,,$ \\
 $r_{1d}=c_{1d}+2 c_{2d} S_0+\ldots\,,$ & $r_{1m}=c_{1m}+2c_{2m} S_0+\ldots\,,$ \\
 $r_{2d}=c_{2d}+\ldots\,,$ & $r_{2m}=c_{2m}+\ldots\,.$ 
\end{tabular}
}
In the previous expression all the terms explicitly depicted are ${\cal O}(1)$ quantities and ellipsis stand for subleading contributions $c_{nx}S^{(n-1)}_0 \sim (\Ff/\Lambda_{\chi})^{n-2}$, for $n>2$ ($x=d,m$). 

There is a subtle point that must be addressed before going on: for generic values of the LECs the shift (\ref{shif}) breaks chiral symmetry. This is most apparent if we lift the scalar and pseudoscalar sources from its vacuum values to arbitrary ones. Namely, if the original scalar field in (\ref{ss}) is a singlet under chiral symmetry, the scalar field after the shift (\ref{shif}) is not. This is so for any value of the parameters, except for those that fulfill
\be
\lambda_1=\frac{c_{1m} {\mathaccent 23 m}_{S}^2 \Ff}{2 c_{2m}}\, .
\ee
If we choose $\lambda_1$ as above, the shift becomes independent of the quark masses ($S_0=-c_{1m}/2c_{2m}$), and hence the scalar field after the shift is still a scalar under chiral symmetry, as it should. Moreover, for this choice, $r_{1m}=0$, and $r_{0m}$ and $r_{0d}$ can be set to $1$ by a redefinition of $B$ and $\Ff$ respectively. The net result is equivalent to choosing $\lambda_1=c_{1m}=0$ in (\ref{ss}) and (\ref{sigmapion}). This has in fact  a simple interpretation. If we impose to our original scalar field to be a singlet under chiral symmetry for any value of the external sources and not mix with the vacuum, then the only solution at tree level is $\lambda_1=c_{1m}=0$. We shall adopt this option from now on. At higher orders these two parameters must be tuned so that no mixing with the vacuum occurs at any given order. Note finally that the value $c_{1m}=0$ is incompatible with the linear sigma model one $c_{1m}=1/4$.

\subsection{Next--to--leading Lagrangian}

In computing loop graphs, we will encounter ultraviolet divergences. These will be regularized within the same dimensional regularization scheme as used in \cite{Gasser:1983yg}, and the elimination of the divergences proceeds through suitable counter--terms. Like in \cite{Gasser:1983yg} we will deal with contributions up to including terms of order $p^4$. Since the singlet fields will only enter in internal propagators, the counter--term Lagrangian we need only involves pions, and hence it has the same functional form as the one in $\chi$PT. The coefficients, however, receive extra contributions due to the appearance of the $c_{ix}$ ($i=1,2$, $x=d,m$) bare parameters. In addition, unlike standard $\chi$PT, now $F$ and $B$ need to be renormalized. In order to take this into account we chose to include explicitly the corresponding counter--terms below. Using the $SU_L(2) \times SU_R(2)$ formalism \cite{Knecht:1997jw}, rather than the $O(4)$ one \cite{Gasser:1983yg}, we have
\bea
\label{lagp4}
{\cal L}^{(4)} = && \frac{1}{4}
\ell_1 \langle D_\mu U D^\mu U^\dagger\rangle^2+\frac{1}{4}
\ell_2 \langle D_\nu U D^\mu U^\dagger\rangle \langle D^\nu U D_\mu U^\dagger\rangle\nn\\
&&+\frac{1}{16}
\ell_3  \langle \chi^\dagger U + \chi U^\dagger\rangle^2+\frac{1}{4}
\ell_4   \langle D^\mu U^\dagger D_\mu \chi +  D^\mu \chi^\dagger D_\mu U\rangle
\\ \nonumber &&
+ Z_1 {\mathaccent 23 m}_{S}^2  \langle \chi^\dagger U + \chi U^\dagger\rangle + 
Z_2 {\mathaccent 23 m}_{S}^2 \langle D_\mu U D^\mu U^\dagger\rangle+\ldots\,. 
\eea
In order to avoid confusion with the values that the LEC take in $\chi$PT and $\chi$PT$_S$, we shall denote the former $l_i$, $i=1,\dots ,4$ and the latter $\ell_i$, $i=1,\dots ,4$. The relations between $l_i$, $l_i^r$ and ${\bar l}_i$, $i=1,\dots ,4$ that appear in the paper are the standard ones in $\chi$PT \cite{Gasser:1983yg}. In this work $\ell_1$ and $\ell_2$ will not be necessary for renormalization. For the observables that we will consider,
the pole at $d=4$ is removed by the following two kinds of renormalization constants which occur in the Lagrangian $ {\cal L}^{(4)}$
\be
\label{l}
\begin{array}{c}
\ell_i := \ell_i^r + \gamma_i \lambda \quad (i=3,4) \,,\quad 
Z_j := Z_j^r + \Gamma_j \lambda \quad (j=1\,,2)\,,
\end{array}
\ee
$$
{\rm with}\quad  \lambda=\frac{1}{16 \pi^2}\left( \frac{1}{d-4} -\frac{1}{2} [ \ln 4\pi + \Gamma^\prime(1)+1]\right)\,.
$$
The first ones, $\gamma_i$, are a simple redefinition of the divergent part in the standard monomials of $\chi$PT 
\be
\label{l3t}
\gamma_3= -\frac{1}{2}+32c_{2m}\left(c_{2m}-c_{2d}\right)-8c^2_{1d}\left(1-4c_{2m}\right)\,,
\quad
\gamma_4=2+4c^2_{1d}\left(1-8c_{2m}\right)+32c_{2d}c_{2m}\,.
\ee
While the second, absent in $\chi$PT, are entirely due to the interaction of pions with the singlet field. In our case, this contribution has two sources; one proportional to $\B$ and other to $\Ff$
\bea
 \Gamma_1 =-2\left(c^2_{1d}-c_{2d}+c_{2m}\right)\,,\quad
 \Gamma_2 =c_{1d}^2-c_{2d}\,.
\eea
Recall that, like $F$ and $B$, $\ell_3^r$ and $\ell_4^r$ are now parameters of a theory different from $\chi$PT and,  hence, their values are expected to differ from the ones in the latter. Note also that our approach differs of that in \cite{Ecker:1988te} in the respect that pions and the singlet field are both dynamical in the same energy range and thus both will be allowed to run inside loops. In that respect, any estimate of the LEC by matching the observables derived from $\chi$PT$_S$ with those obtained from a Lagrangian of resonance exchange as in \cite{Ecker:1988te} should keep the singlet field $S$ in the latter as a dynamical low--energy degree of freedom. Loop effects of scalar resonances coupled to pseudo--Goldstone bosons have been studied in \cite{Rosell:2005ai,Rosell:2006dt,Portoles:2006nr,Rosell:2004mn}.

\section{The axial--vector two--point function}\label{ax2pf}

\begin{figure}
\centerline{
\begin{tabular}{cccc}
\includegraphics[width=3cm]{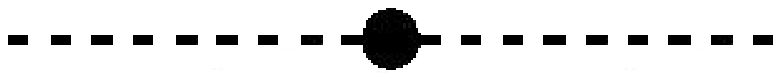}& \includegraphics[width=3cm]{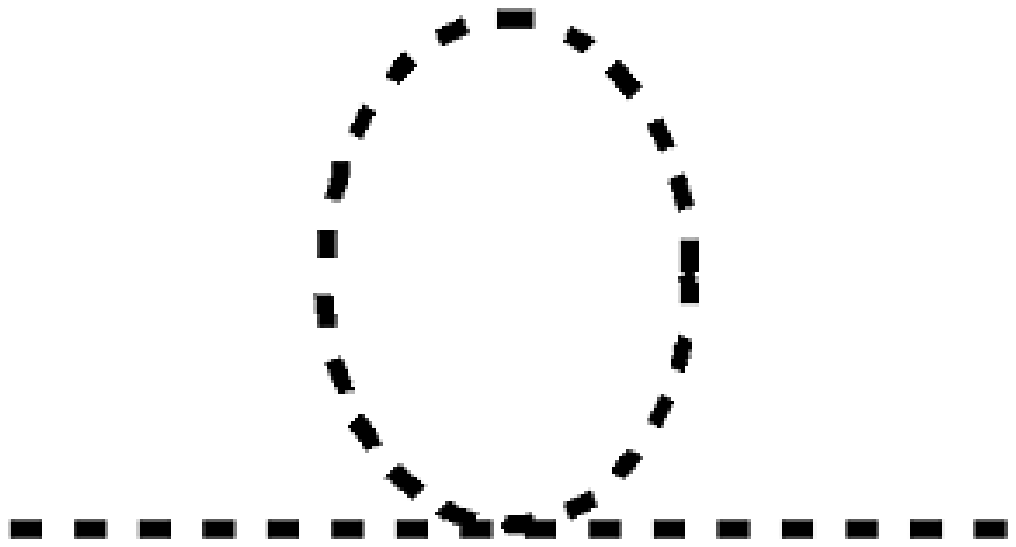}& \includegraphics[width=3cm]{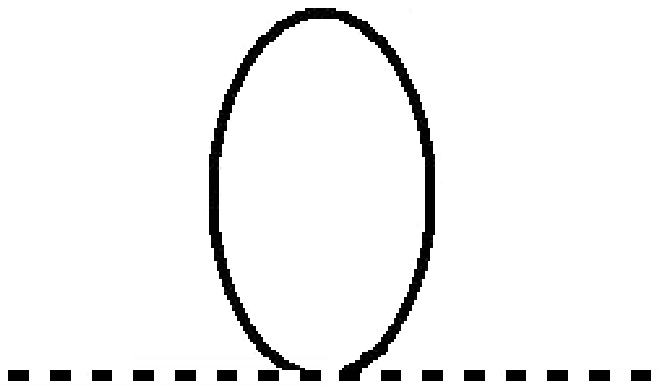} & \includegraphics[width=3cm]{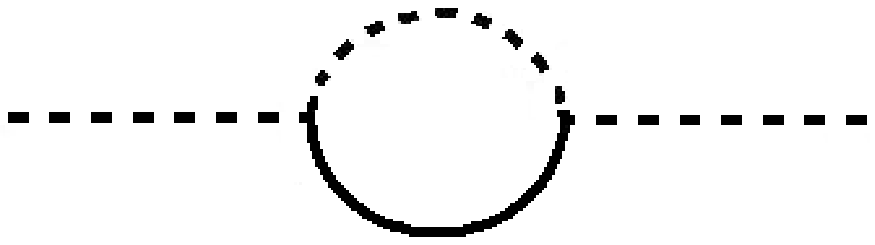} \\
(a) & (b) & (c) & (d) \\
\end{tabular}
}
\caption{Self energy diagrams. Solid lines denote the scalar and the dotted ones pions. Diagram (a) corresponds to NLO counterterms}
\label{auto}
\end{figure}

\begin{figure}
\centerline{
\begin{tabular}{cccc}
\includegraphics[width=3.3cm]{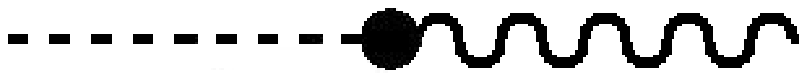} & \includegraphics[width=3cm]{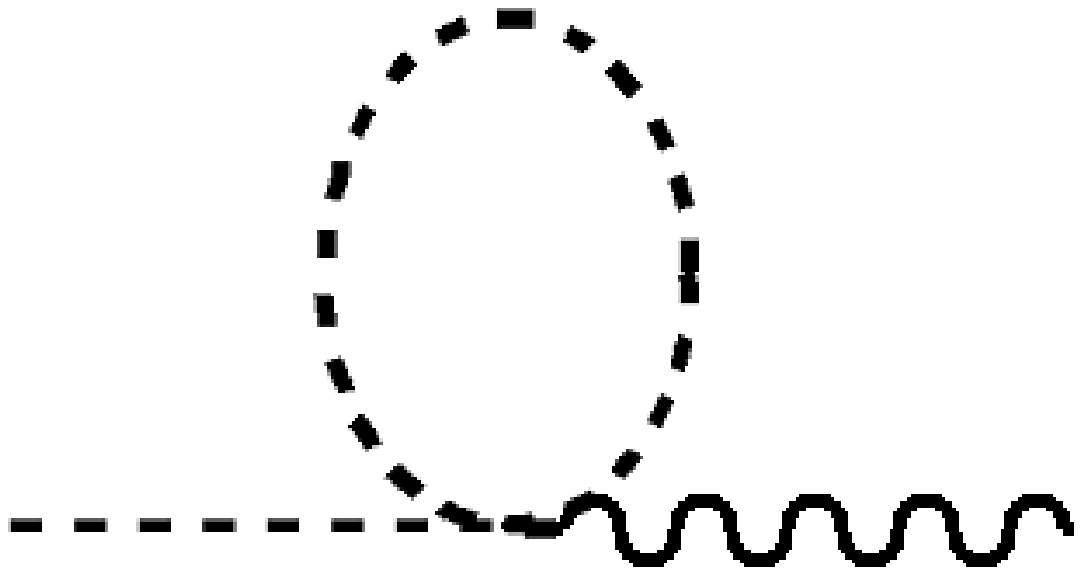} & \includegraphics[width=2.8cm]{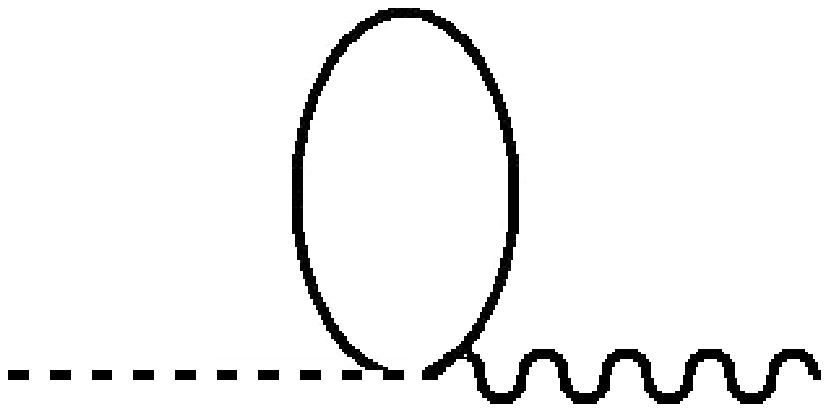} &\includegraphics[width=3.3cm]{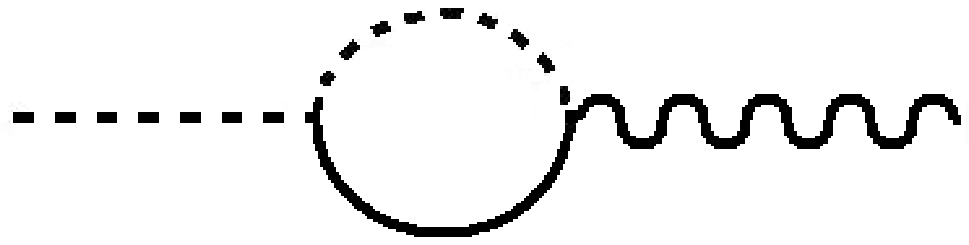} \\
(a) & (b) & (c) & (d) \\
\end{tabular}
}
\caption{Diagrams contributing to the decay constant. Solid lines denote the scalar and the dotted ones pions. 
Diagram (a) corresponds to NLO counterterms.}
\label{decaypi}
\end{figure}

We are now in the position to perform a complete NLO analysis of the pion mass and decay constant, including the 
radiative correction due to the singlet field. In order to calculate them we focus in the sequel in the axial--vector two--point Green function. The first quantity will be defined through the position of the pole in the two--point function while the second one can be obtained directly from its residue.

At the Born level the expressions for the pion mass and decay constant do not differ from those of the standard $\chi$PT 
theory while at NLO corrections are slightly more cumbersome. We denote by $m^2_{\rm PS}$ and $F_{\rm PS}$ the pion mass and 
the pion decay constant respectively calculated at NLO, whereas we keep $m_\pi=2B\hat m$ and $F$ for the same quantities at LO. 
The diagrams contributing at ${\cal O}(p^4)$ to $m^2_{\rm PS}$ and $F_{\rm PS}$ are represented in Fig. \ref{auto} and in Fig. \ref{decaypi} respectively. In both figures the diagram (a) is the usual couterterm contribution, and the diagram (b) the usual tadpole contribution, 
already encounter in the standard theory. Diagrams (c) and (d) are new and appear because of the singlet field. 

We cast the expressions for the mass and decay constant as
\eqn{fmpi}{
m_{\rm PS}^2=&
 2 B \hat{m}+ U_m + P_m + {\cal O}(p^6)\,,\cr
F_{\rm PS} = & F\left(1+U_F + P_F +  {\cal O}(p^6)\right)\,.
}
They contain contributions related to the unitarity cut ($U_m$, $U_F$) in the s--channel, Fig. \ref{auto}(d) and Fig. \ref{decaypi}(d), and a polynomial term in $s\,\,(=m_\pi^2)$ which includes logarithms ($P_m$, $P_F$).

Notice that, as we introduced a new degree of freedom, the infra--red physics is modified and in particular the analytic structure of the amplitudes. While $\chi$PT finds its first cut at NNLO for the pseudoscalar two--point function, located at $3 m_\pi$, $\chi$PT$_S$ has already at NLO a cut located at $m_S+m_\pi$. This enhancement can be understood if one chooses to interpret the scalar field as a simulation of a strong two--pion rescattering. Consider for example the three pion contribution to the pion self--energy. If we take the two--pion subdiagram, say $t^{(4)}$, and replace it according to $t^{(4)}\rightarrow (t^{(2)})^2/\left(t^{(2)}-t^{(4)}\right)-t^{(2)}$, as suggested by the Inverse Amplitude Method (IAM), the first term would correspond to the pion--sigma contribution in $\chi$PT$_S$ and the second one to a pion tadpole, both contributions being of $\mathcal{O}\left(p^4\right)$ in the pion self--energy in $\chi$PT$_S$.

The expressions for the mass are given by
\eqn{Up}{
U_m& =-\frac{4c^2_{1d}}{F^2} \bar{J}(m_\pi^2,m_{S}^2;m_\pi^2) \left(m^2_S-2m^2_{\p}\right)^2\,,\cr
P_m& = \frac{4m^4_{\p}}{F^2}\left(\frac{\m_S-\m_{\p}}{\D_{\p S}}\right)\left(c^2_{1d}m^2_{S}-4c_{2m}\G_1\D_{\p S}\right) 
+\frac{m^4_{\p}}{16\p^2F^2}\gamma_3\bar{\ell}_3 +\frac{m^2_\p {\mathaccent 23 m}_{S}^2}{8\p^2F^2} \G_1\bar{Z}_1 \,,}
and for the decay constant
\eqn{UF}{
U_F = &\frac{2c^2_{1d}}{F^2 m^2_{\p}} \bar{J}(m_\pi^2,m_{S}^2;m_\pi^2)\left(\frac{2m^2_{\p}-m^2_S}{4m^2_{\p}-m^2_S} \right) \left(14m^4_{\p}-15m^2_{\p}m^2_S+3m^4_S\right)\,,\cr
P_F = &\frac{c^2_{1d}}{8\p^2 F^2}\frac{\left(m^2_S-2m^2_{\p}\right)^2}{4m^2_{\p}-m^2_S}+\frac{4m^2_{\p}}{F^2}\left(\frac{\m_{\p}-\m_S}{\D_{\p S}}\right) \left(\frac{c^2_{1d}\left(m^2_S-2m^2_{\p}\right)^2}{\left(4m^2_{\p}-m^2_S\right)}+4c_{2m}\G_2\D_{\p S}\right)  \cr
&+ \frac{m^2_{\p}}{32 \p^2 F^2}\gamma_4 \bar{\ell}_4+\frac{ {\mathaccent 23 m}_S^2}{8\p^2F^2}\G_2\bar{Z}_2 \,.}

The functions $\bar{J}$ and $\mu_a$ ($a=\pi\,,S$) are displayed in the Appendix. In addition we have used the scale independent quantities $\bar{\ell}_i$ and $\bar{Z}_j$, that are defined as follows:
\bea
\label{l2}
\ell_i^r= \frac{ \gamma_i}{32\pi^2}\left[\bar{\ell}_i+\ln \left( \frac{m_\pi^2}{\Lambda^2}\right) \right]\, (i=3,4)\,,
\quad
Z_j^r= \frac{\Gamma_j}{32\pi^2}\left[\bar{Z}_j+\ln \left(\frac{m_{S}^2}{\Lambda^2}\right) \right]\, (j=1,2)\,,
\eea
where $\Lambda$ is the renormalization scale.

Both quantities in (\ref{fmpi}) have the following virtues that constitute non--trivial tests on their correctness: 
\begin{enumerate}

\item Despite their appearance, they are finite in the chiral limit, $\hat{m}\to 0$. More explicitly, in this limit the pion mass vanishes, as it should, while the decay constant reads
\be
F_{\rm PS}=F\left(1+\frac{{\mathaccent 23 m}_S^2}{8 F^2 \p ^2}\left[\bar{Z}_2 \G_2+\frac{1}{2}c_{1d}^2\right]\right)\,.
\label{flimit}
\ee

\item Setting $c_{ix} \to 0\, (x=m,d)$ in 
(\ref{fmpi}) they reduce to their standard $\chi$PT values
\eqn{mpis}{
m_{\rm PS\,\,{\rm \chi PT}}^2= 
 2 B \hat{m} \left(1- \frac{1}{16 \pi^2 F^2}  B \hat{m} \bar{l}_3\right)
\,,\quad 
F_{\rm PS\,\,{\rm \chi PT}} =  F \left( 1 +  \frac{1}{8\pi^2  F^2}  B \hat{m} \bar{l}_4 \right)\,.
}
\end{enumerate}

To conclude this section we integrate out the singlet field. In the infrared limit, $m_\pi^2 \sim p^2 \ll {\mathaccent 23 m}_S^2$,  $\chi$PT$_S$ has to reduce to $\chi$PT, where the only dynamical degrees of freedom are the pions \cite{Gasser:1983yg}. In oder to do so we keep ${\mathaccent 23 m}_S^2$ fixed and expand the above observables around $m_{\p} \sim 0$. At NLO order in this expansion we indeed recover (\ref{mpis}) after the identification
\be
\begin{split}
\bar{l}_3=&-2\bar{\ell}_3\g_3- 32\G_1 \ln\left(\frac{m^2_{\p}}{{\mathaccent 23 m}_S^2}\right)c_{2m}-\frac{4}{3}c_{1d}^2 \left(1+12c_{2m}\right)\,,\\
\bar{l}_4=&\frac{1}{2}\bar{\ell}_4\g_4-16\G_2 \ln\left(\frac{m^2_{\p}}{{\mathaccent 23 m}_S^2}\right) c_{2m}+c_{1d}^2 \left[1-8c_{2m}+2\ln\left(\frac{m^2_{\p}}{{\mathaccent 23 m}_S^2}\right)\right]\,.
\end{split}
\label{decl3}
\ee

\section{The scalar field two--point function}\label{s2p}

\begin{figure}
\centerline{
\begin{tabular}{ccc}
\includegraphics[width=2.5cm]{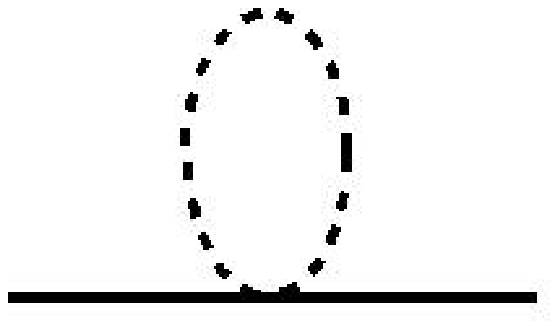}& \includegraphics[width=2.5cm]{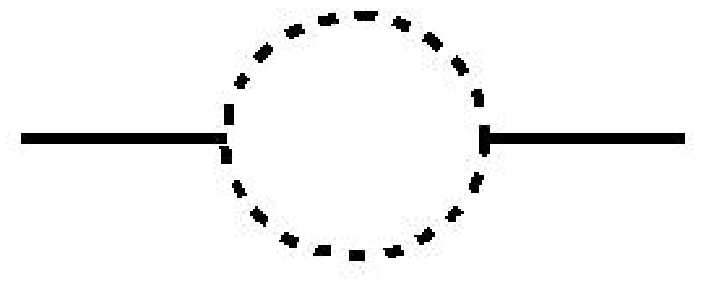}& \includegraphics[width=2.5cm]{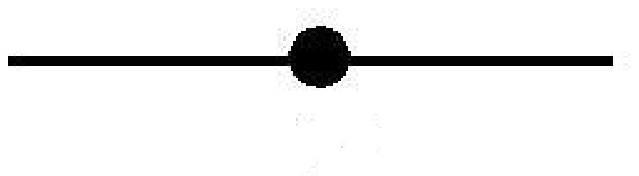} \\
(a) & (b) & (c)  \\
\end{tabular}
}
\caption{Diagrams contributing to the scalar field self--energy. Solid lines denote the scalar and the dotted ones pions.}\label{scselfe}
\end{figure}

In order to calculate the scalar field two--point function at NLO, we need to enlarge our initial set of operators in (\ref{lagp4}) to provide the needed counter--terms to renormalize it
\be
\begin{split}
\delta {\cal L} =& f_{2p} \Box S  \Box S  + d_{2m} \partial_\mu S \partial^\mu S\langle \chi^\dagger U + \chi U^\dagger\rangle + b_{2m}S^2 \langle \chi^\dagger U + \chi U^\dagger\rangle^2 + a_{2m}S^2\langle \chi^\dagger \chi\rangle \\
&+ e_{2m}S^2\Re[ \det(\chi)] \,.
\end{split}
\label{lp4scalar}
\ee
Notice that the first of these counter--terms may be eliminated by using the LO equation of motion. We keep it for convenience, it renormalizes contributions to a contact term, to the scalar wave function, and to the scalar mass in the two--point function. The divergent parts of the low--energy constants in (\ref{lp4scalar}) are determined by the cancellation of divergences in the two--point function. The combination $d_{1m}\equiv 32 b_{2m}+4 a_{2m}+2 e_{2m}$ is indistinguishable in this quantity, and is renormalized as a whole. The divergent parts read as follows
\be
f_{2p}:=f^r_{2p}+\Gamma_f \lambda\,,\quad d_{im} := d_{im}^r + \D_i \lambda\,,\,(i=1\,,2)\,,
\ee
with
\be
\Gamma_f=\frac{12 c^2_{1d}}{F^2}\,,\quad  \D_{1}=\frac{24}{F^2}\left(c_{2m}-c_{2d}+6c^2_{1d}\right)\,,\quad \D_{2}=-\frac{9c^2_{1d}}{F^2}\,,
\ee
and $\lambda$ is defined as in (\ref{l}). The set of diagrams contributing at NLO to the two--point function is shown in Fig. \ref{scselfe}. The contributions at NLO to the off--mass--shell two--point function can be disentangled as 
\eqn{s2point}{ 
\Sigma_{2p\text{\,NLO}}=U_S + P_S 
\,,}
where, as in pion two--point function, we find a contribution related to the unitarity cut, $U_S$, and a polynomial term in $s$ including the logarithm term $P_S$. Explicitly
\eqn{Us}{
U_S& =\frac{6c^2_{1d}}{F^2} \bar{J}(m_\pi^2,m_{\pi}^2;s) \left(s-2m^2_{\p}\right)^2\,,\cr
P_S& =\frac{s^2}{32\p^2}\Gamma_f\bar{f}_{2p}+\frac{m^4_{\p}}{32\p^2}\D_1\bar{d}_{1m}+\frac{m^2_{\p}s}{4\p^2}\D_2\bar{d}_{2m}\,,}
where we have used the scale independent quantities $\bar{d}_{im}$
\be
f^r_{2p}=\frac{\Gamma_f}{32\p^2}\left[\bar{f}_{2p}+\ln\left(\frac{m^2_{\p}}{\L^2}\right)\right]\,,\quad d^r_{im}=\frac{\D_i}{32\p^2}\left[\bar{d}_{im}+\ln\left(\frac{m^2_{\p}}{\L^2}\right)\right]\,,\,(i=1,2)\,.
\ee
The scalar mass at NLO, defined as the pole of the scalar field two--point function, reads as follows
\be
\begin{split}
m^2_{S,\,{\rm NLO}}=& m^2_S-\frac{m^4_S}{32\p^2}\Gamma_f\bar{f}_{2p}-\frac{m^4_{\p}}{32\p^2}\D_1\bar{d}_{1m}-\frac{m^2_{\p}m^2_S}{4\p^2}\D_2\bar{d}_{2m}\\
&-\frac{6c^2_{1d}}{F^2} \bar{J}(m_\pi^2,m_{\pi}^2;m^2_S) \left(m^2_S-2m^2_{\p}\right)^2\,.
\end{split}
\label{ms}
\ee

\begin{figure}
  \centerline{\includegraphics[width=7cm]{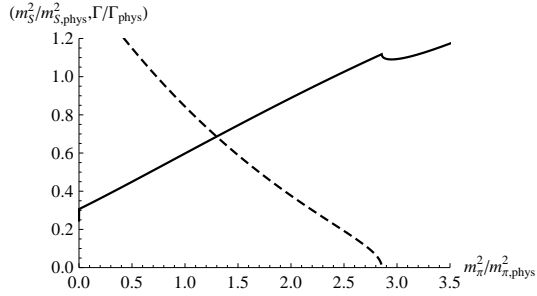}}
\caption{Plot of the scalar mass (solid line) up to NLO and the sigma decay width (dashed line) at LO both normalized to their respective physical values as a function of the quark mass. $f^r_{2p}$ and $d^r_{im}$ are set to zero, $c^2_{1d}$ is taken from (\ref{cc1d}), and $c_{2d}$ and $c_{2m}$ from (\ref{nlocpts}).}  \label{msvsmpi}
\end{figure}
Contrariwise to \cite{Hanhart:2008mx} we do not find branching points in the behavior of (\ref{ms}) when the pion mass increases. In \cite{Hanhart:2008mx} two conjugate poles in $s$ of the amplitude are found. The branching is a result of the different behaviour of these poles when they reach the real axis. In $\chi$PT$_S$ the denominator in the amplitude has a second order polynomial in $s$ like in the IAM. The non-analytic piece has the same structure as the s-channel contribution to the IAM, but the t- and u-channel contributions are absend. They are treated as perturbations in $\chi$PT$_S$. If we look for poles we find two complex conjugate poles like in \cite{Hanhart:2008mx}. However when we take into account the counting, we observe that the self-energy needs to be resummed when $s-m_S^2 \sim m_S^4/\Lambda_\chi^2$ only. As a consequence, only one pole remains, now with an imaginary part, related to the sigma decay width.

If we set to zero the finite part of all the unknown counter--terms ($f^r_{2p}=d^r_{im}=0$), the scalar mass (\ref{ms}) scales almost linearly with the pion mass up to the threshold region, see full curve in Fig. \ref{msvsmpi}, where the $\pi$--$\pi$ production channel closes.

The scalar decay width can be read from the unitary part $U_S$ once we plug the logarithm dependence of the $\bar{J}$ function and set it to be on--mass--shell
\be
\frac{\G}{2}=\frac{3c^2_{1d}}{8\p F^2 m_{S}}\sqrt{1-\frac{4m^2_{\p}}{m^2_{S}}}\left(m^2_S-2m^2_{\p}\right)^2\,.
\label{gamma}
\ee
Notice that it only depends on a single unknown LEC, $c_{1d}$. Using the standard values for $F \sim F_\pi$ and 
$ m_{\p}$,  and taking specific values for the mass and width of the sigma resonance from \cite{Caprini:2005zr}
\be
\label{inpp}
F_{\p}=92.419\,\rm{MeV}\,,\quad  m_{\p}=139.57\,{\rm MeV}\,, \quad m_{S,\,\,\rm CCL}=441.2\,{\rm MeV}\,, \quad\G_{\rm CCL}/2=272\,\rm{MeV}\,,
\ee
we obtain from (\ref{gamma})
\be
\label{cc1d}
c^2_{1d}=0.457\,.
\ee

\section{Matching with lattice data: the pion mass and decay constant}\label{matchlatdat}

The expressions for the pion mass and decay constant (\ref{fmpi}) depend on several LEC not constrained by chiral 
symmetry,  $\ell^r_3\,, \ell^r_4\,,c_{1d}\,, c_{2d}\,, c_{2m}$ in addition to the quark masses and the bare parameters $F$, $B$, ${\mathaccent 23 m}^2_{S}$. At this point, and for fitting purposes, the finite part of the counterterms $Z_{1,2}^r$ can be absorbed into $F$, $B$. Then, we have eight independent parameters at our disposal at NLO. 

Lattice QCD offers a new arena for determining the LEC. Unlike physical experiments, lattice calculations use different unphysical quark masses, providing for each point what can be considered as an uncorrelated {\sl experimental} datum with Gaussian errors. We will use the lattice data based on maximally $n_f=2$ twisted fermions to fit the LEC \cite{Baron:2009wt}. More precisely the data ensembles labeled as A1-- A4, B1-- B6, C1-- C4, D1 and D2 in the Appendix C of that reference. Both finite volume effects and discretization errors are small in the data sets we use, and will be ignored in the following\footnote{This also implies that the isospin violation due to the twisted mass term in the lattice lagrangian is small, as it vanishes in the continuum limit, see \cite{Dimopoulos:2009qv}. Lattice data in ref.\cite{Baron:2009wt} corresponds to charged pions.}. We will also use in our analysis a single lattice scale $r_0=0.446\, {\rm fm}$ for all data sets as a simplification, which is justified because its value varies very little from one data set to another.

Given the limited quantity and quality of the available data, the number of free parameters is too large to expect a brute force best fit to provide sensible values for all of them. We have rather used a general three--fold strategy:

\begin{enumerate}

\item We identified $F_{\rm PS}$ (\ref{fmpi}) with its physical value at the physical pion mass (\ref{inpp}). This allows to write $F$ as a function of the remaining parameters. This determination is done perturbatively. 

We have used the same procedure to fix the bare scalar mass through (\ref{spredef}): one imposes to the tree level scalar mass to take its physical value (\ref{fmpi}). 

\item The $c_{1d}$ parameter appears in our expressions always squared, hence we will use $c^2_{1d}$ as the free parameter. As we have seen its value can be extracted through the sigma decay width, (\ref{cc1d}). 

\item For each point in the $c_{2x}$ ($x=d,m$) parameter--space we fitted the lattice data minimizing an augmented chi--square distribution that includes both observables in (\ref{fmpi}) \cite{Schindler:2008fh}. The augmented chi--square distribution is defined as the sum of the chi--square functions for each observable together with a set of priors for each of the free parameters to be fitted
\be
\begin{split}
\chi^2_{aug}=\chi^2_{m^2_{PS}}& + \chi^2_{F_{\rm PS}}+\chi^2_{prior}\,, \\
\chi^2_{g}=\frac{1}{n}\sum\limits^n_i \frac{\left(g(\hat{m}_i)-g_i\right)^2 }{\d^2_{g_i}} \quad ,& \quad \chi^2_{prior}=\frac{1}{N}\sum\limits^N_i \frac{\left(\ln x_i -\ln a_i\right)^2 }{\ln^2 R_i}\,, 
\end{split}
\ee
where $g$ stands either for $m_{PS}$ or $F_{PS}$ and $g_i$ for the corresponding lattice data at the quark mass $\hat{m}_i$. Furthermore, $x_i$ refer to the fitted parameters $N$ being their total number. At NLO $N=3$ and $x_i=\{B/(2600\, {\rm MeV})\,,\ell^r_3\,,\ell^r_4\}$. The prior information on the LEC is obtained from naive dimensional analysis. For instance, if the parameter $x_k$ is of order ${\cal O}(1)$, we expect it to be in the range $0.1<\|x_k\|<10$, which translates  to setting $\ln(\|a_k\|)=0$ and $\ln(R_k)=1$ for the \textit{k}th parameter. We have taken logarithms in the prior functions to achieve equal weights for the subranges $0.1<\|x_k\|<1$ and $1<\|x_k\|<10$.
\end{enumerate}

Within the previous outlined procedure we have fitted the expressions for the observables $m_{\rm PS}^2$ and $F_{\rm PS}$, (\ref{fmpi}), 
to the full range of available quark masses $m_q$, the $\overline{MS}$ running mass at the scale $\mu = 2$GeV. Throughout this section we have used $\Lambda=770\,{\rm MeV}$ as the renormalization scale value. The chi--square per degree of freedom, $\chi^2_{d.o.f}$ is 
\be
\chi^2_{d.o.f}=\frac{n\chi^2_{m^2_{PS}}+n\chi^2_{F_{PS}}}{2n-1-N'}\,,
\ee
where $N'$ is the number of free parameters, including the fitted and scanned ones. Thus for the NLO fit $N'=5$.

\subsection{$\chi$PT results}

Before we move ahead with the $\chi$PT$_S$ case, and for comparison purposes, we present the outputs we obtain for $\chi$PT. From the previous steps in the fitting procedure we only make use of the first to fix the physical point of $F_{\rm PS}$ and the third to introduce the priors to the corresponding parameters.

\subsubsection{Born approximation}

At LO, the bare parameter $F$ is fixed by the physical decay constant $F=92.419\,{\rm MeV}\,$ and only $B$ is a free parameter, obtaining
\bea
B=  2250.4\,{\rm MeV}\,,
\eea
with $\chi^2_{d.o.f}=\frac{1560}{30}$. 

\subsubsection{Next--to--leading order results}

For $\chi$PT the NLO expressions in (\ref{fmpi}) have three free parameters, $B$, $l^r_3$, and $l^r_4$ while $F$ has been fixed perturbatively at the physical point. The fitting procedure leads to
\bea
\label{lss}
B=2499(10)\,{\rm MeV}\,,\quad
l^r_3=0.91(6)\times10^{-3}\,,\quad
l^r_4=7.13(5)\times10^{-3}\,.
\eea
Using these values on the constraints imposed in the fist step of the fitting procedure we obtain $F$
\be
F=86.36(1)\,{\rm MeV}\,,
\ee
with the final value $\chi^2_{d.o.f}=\frac{16.9}{28}\,.$

The results, together with the lattice data, are plotted in dashed lines in Fig. \ref{plotoneloop}. The adjustment for the pion mass to the lattice points is quite remarkable, with only a small deviation for large values of the quark masses. The pion decay constant fit also reaches a good agreement with the lattice data.

The results for the LEC, (\ref{lss}), are compatible with standard values in the literature
\be
\begin{tabular}{clcl}
 \cite{Colangelo:2010et}& $\bar{l}_3=3.2\pm 0.8$\,,&  \cite{Colangelo:2001df}&  $\bar{l}_4=4.4\pm 0.2$\,. \\ 
 \scriptsize{eq} (\ref{lss}) & $\bar{l}_3=3.99\pm 0.04 $\,,& \scriptsize{eq} (\ref{lss}) &  $\bar{l}_4=4.54\pm 0.01$\,. \\
\end{tabular}
\label{standardOLD}
\ee
These estimates are in reasonable agreement with those obtained by resonance saturation \cite{Ecker:1988te}
\be
\begin{tabular}{ll}
 $\bar{l}_3=2.9\pm 2.4$\,,  &  $\bar{l}_4=4.3\pm 0.9$\,. \\ 
\end{tabular}
\ee

The determination of the uncertainties in this section has been performed as follows. We assume each data point corresponds to Gaussian distribution with expected value and variance defined by the data point value and uncertainty respectively, then we generate random data sets according to these distributions and perform a fit for each one. The final parameters are obtained from the average of the results of these fits, while the uncertainty is obtained from the variance.
Comparing our results with those from Table 1 in \cite{Baron:2009wt}, $B=2638(149)(132){\rm MeV}$, $F=85.91(7)(^{+78}_{-7})$, $\overline{l}_3=3.50(9)(^{+9}_{-30})$, $ \overline{l}_4= 4.66(4)(^{+4}_{-33})$, we observe that $F$, $\bar{l}_4$ and $B$ values are within one sigma while $\bar{l}_3$ is within two sigmas. Note that our uncertainty analysis does not include systematic uncertainties because we have used only one set of data ensembles and we do not include finite size corrections. Statistical uncertainties in our fit are significantly smaller than those of \cite{Baron:2009wt}. This is because we have taken $r_0$ as a fix value rather than as an additional free parameter. If we estimate the uncertainty of $r_0$ as the one given in \cite{Baron:2009wt} and we extrapolate the effect to our results we obtain uncertainties in the same range as in \cite{Baron:2009wt}. Furthermore, since $m^2_{\rm PS}$ depends quadratically on $r_0$ while $F_{\rm PS}$ only linearly, our $F$ and $\bar{l}_4$ should be in better agreement with those of \cite{Baron:2009wt} than our $B$ and $\bar{l}_3$, as it is the case.

The estimation of uncertainties above is not directly applicable to the following section because for the fit to $\chi$ PT$_S$ expressions some of the parameters will be obtained by scanning a suitable range. In any case, we are not interested at this point in an accurate determination of the $\chi$PT$_S$ parameters but rather in finding out if parameter sets of this theory exists which are both compatible with lattice data and with physical observables.

\subsection{$\chi$PT$_S$ results}

The LO $\chi$PT$_S$ expressions are identical to those of standard $\chi$PT, therefore the same analysis as in the previous section applies. At NLO appear four extra free parameters to fit $c_{2d}\,,c_{2m}\,,\ell^r_3$ and $\ell^r_4$ and the non--analytical dependence on the light quark masses is greatly augmented (\ref{fmpi}). 

\begin{figure}
  \centerline{
  \includegraphics[width=7cm]{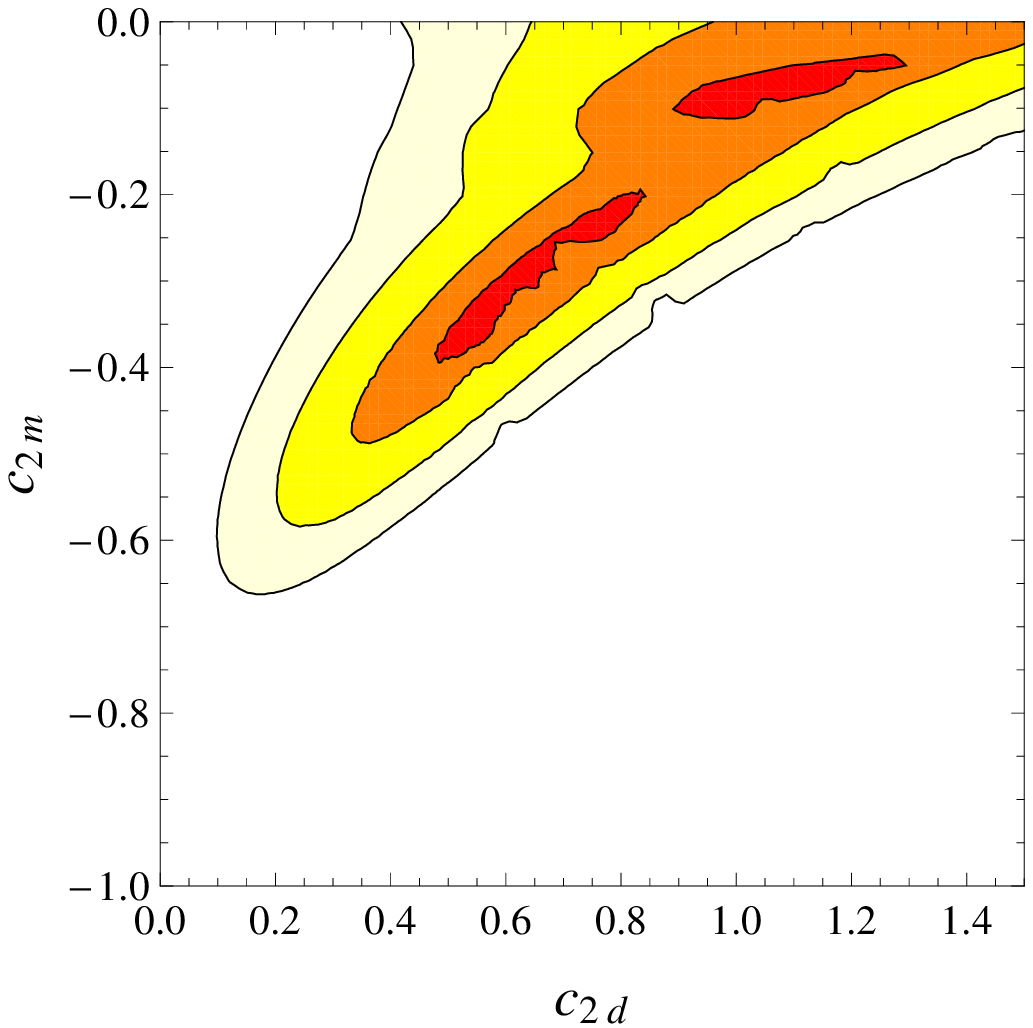}
 \raisebox{2cm}{\includegraphics[width=2cm]{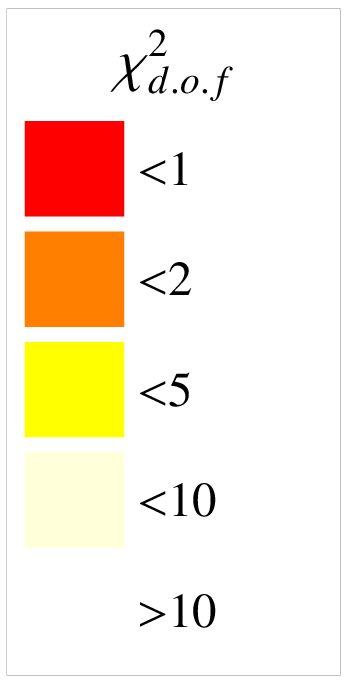}}}
  \caption{$\chi^2_{d.o.f}$ swept over a $(c_{2d},c_{2m})$ grid corresponding to fits to NLO order expressions. The fits are forced to reproduce the pion decay constant, the mass of the sigma resonance and its width at the physical point.}
  \label{plotscan}
\end{figure}
\begin{figure}
  \centerline{
  \includegraphics[width=8cm]{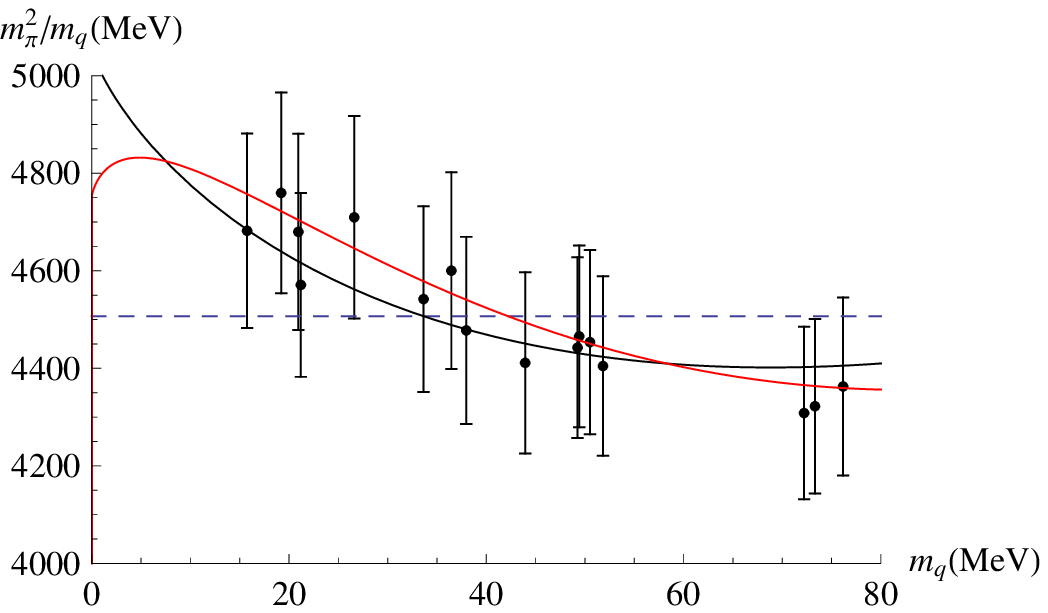}
  \includegraphics[width=8cm]{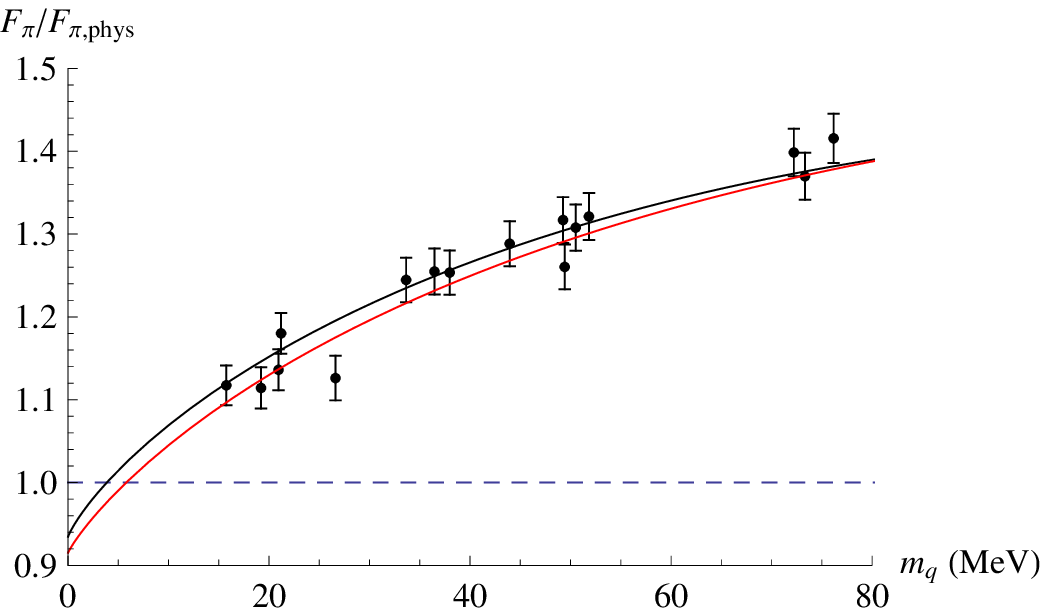}}
  \caption{The best fits of the LO (dashed line), NLO $\chi$PT (black solid line) and $\chi$PT$_S$ (red solid line) expressions. Note that the LO expression is the same for $\chi$PT and $\chi$PT$_S$.}
  \label{plotoneloop}
\end{figure}
The relative large number of free parameters appearing at NLO are an indication that there is no unique solution for the best fit. Indeed, if we look at the contour level plot of the $\chi^2_{d.o.f}$ corresponding to the $(c_{2d},c_{2m})$ region scanned, shown in Fig. \ref{plotscan}, we can see regions of parameter sets with $\chi^2_{d.o.f}$ smaller than one. Thus any parameter set on those regions has to be considered a valid solution. Keeping this in mind, the following are the  results for the best fit obtained, which have been used for Fig. \ref{plotoneloop}

\bea\label{nlocpts}
&&
B= 1680.5\,{\rm MeV}\,,\quad
c_{2d}=1.21\,,\quad
c_{2m}=-0.083\,,\quad\\
&&
\ell^r_3=-1.12\times10^{-3}\, \,,\quad
\ell^r_4=6.94\times10^{-3}\,\,.\nonumber
\eea 
Using these in (\ref{spredef}) and (\ref{fmpi}) we obtain the values of the remaining parameters
\be
F=101.2\,{\rm MeV}\,,\quad
{\mathaccent 23 m}_{S}=426\,{\rm MeV}\,,
\ee 
with 
\eqn{chidisntl}{
\chi^2_{d.o.f}= \frac{16.7}{26}\,.
}

\subsubsection{Scalar resonance contribution to $\chi$PT low--energy constants}

It is instructive to show how the LEC $\ell^r_3$ and $\ell^r_4$ of $\chi$PT$_S$ above compare with the standard LEC of $\chi$PT. This is done through the matching formula (\ref{decl3}). We obtain that the corresponding values of $\bar{l}_3$ and $\bar{l}_4$ read
\be
\bar{l}_3= 3.40\,, \quad\bar{l}_4=5.16\,,
\ee
thus, $\bar{l}_3$ is compatible with the literature values in (\ref{standardOLD}) and $\bar{l}_4$  is somewhat higher. From (\ref{decl3}) we can easily find out the fraction of $\bar{l}_3$ and $\bar{l}_4$ that is exclusively due to the light scalar field by setting $\ell^r_3$ and $\ell^r_4$ to zero. It amounts to a $43\%$ for $\bar{l}_3$ (with opposite sign) and to a $20\%$ for $\bar{l}_4$. This suggest that the impact of the singlet field in both $\bar{l}_3$ and $\bar{l}_4$ is quite substantial. Note that the contributions of the scalar field to these LEC comes entirely through loops, and hence have nothing to do with the tree--level contributions obtained in \cite{Ecker:1988te}.

\subsubsection{Quark mass determination}

The last application we have explored is the determination of the light quark masses, and the comparison with the results obtained from $\chi$PT and lattice QCD. Given a set of parameters the expressions for $m^2_{\rm PS}$ (\ref{fmpi}) and (\ref{mpis}), become a function of $\hat{m}$. Setting $m^2_{\rm PS}$ to the physical value of the pion mass we can solve the equation to obtain the value of $\hat{m}$ at the physical point. The results obtained for $\hat{m}$ for the best $\chi$PT$_S$ and $\chi$PT fits are displayed in Table \ref{qm}. The expressions used for light quark masses match the order at which the fit has been performed, at NLO the equation has been solved perturbatively.

\begin{table}
\begin{center}
\begin{tabular}{|c|c|} \hline
          & $\hat m$ (MeV)\\\hline\hline
Latt.(${\rm stat}$)(${\rm syst}$) &  $3.469(47)(48)$\\ \hline\hline
Beyond NLO $\chi$PT & $3.54(19)(17)$ \\ \hline\hline
$\chi$PT, LO fit & $4.32$\\ \hline
$\chi$PT, NLO fit &  $4.39$ \\ \hline
$\chi$PT$_S$,  NLO fit & $3.26$ \\ \hline
\end{tabular}
\end{center}
\caption{The values obtained for the light quark masses from our fits (bottom panel), at LO $\chi$PT and $\chi$PT$_S$ are 
identical. In the central panel we show the lattice QCD results from Ref. \cite{Baron:2009wt} that use chiral extrapolations
 beyond NLO $\chi PT$ . In the upper panel we show the values from lattice QCD at the physical point \cite{Durr:2010vn}.}
\label{qm}
\end{table}

\section{S--wave $\pi$--$\pi$ scattering lengths}\label{swpipisl}

\begin{figure}[h,b]
\centerline{
\begin{tabular}{cc}
\includegraphics[width=2.2cm]{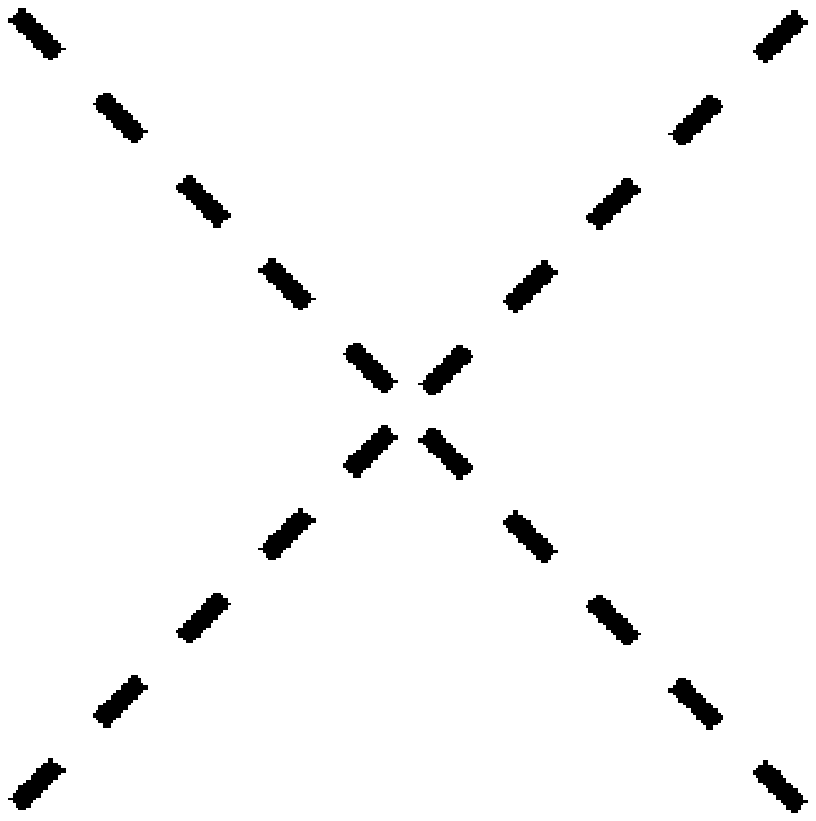}&\hspace{2cm}
  \includegraphics[width=4cm]{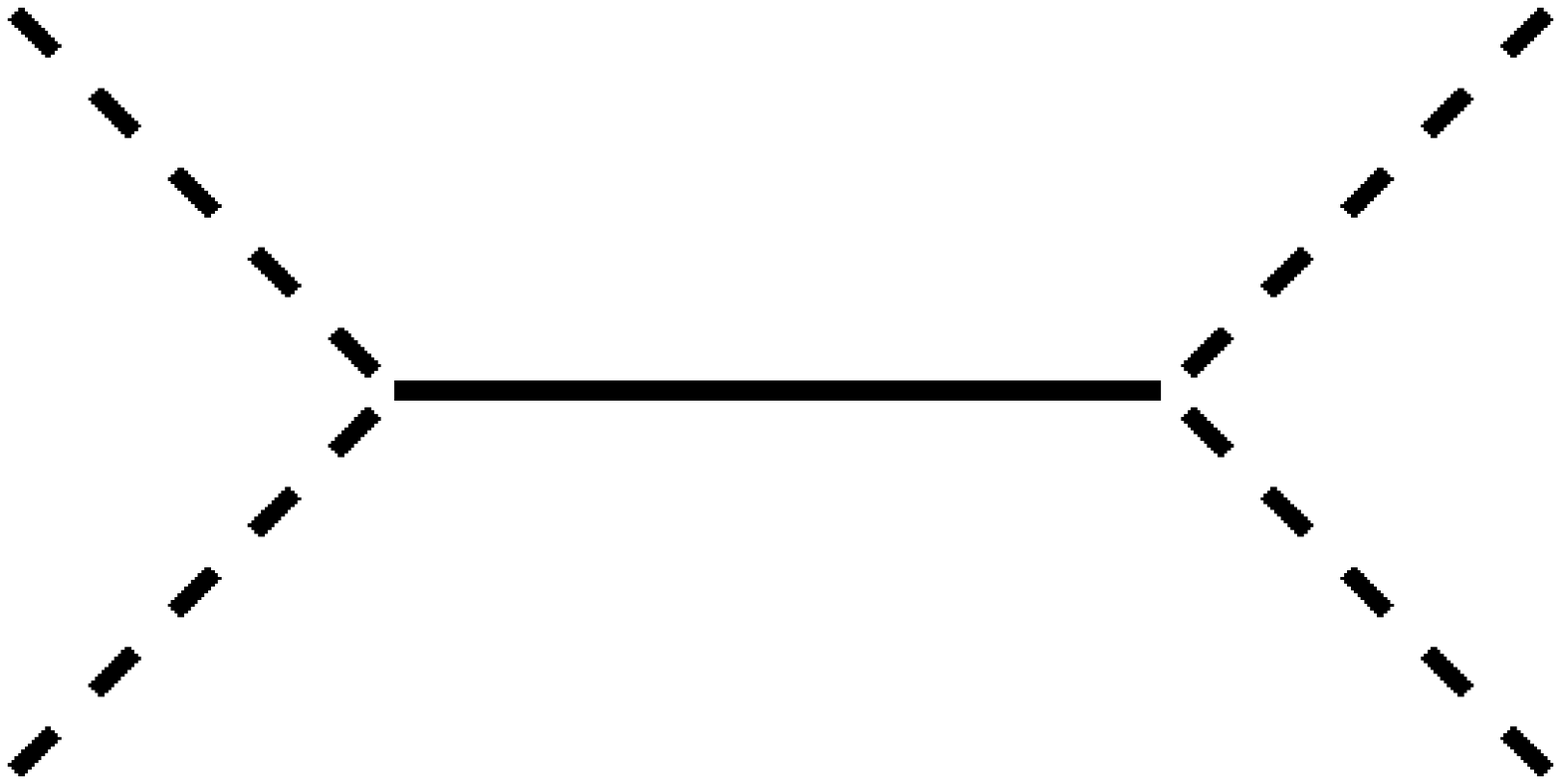}\\
(a) &\hspace{2cm} (b)  \\
\end{tabular}
}
\caption{Diagrams contributing to the scattering lengths. Solid lines denote the scalar and the dotted ones pions.}
\label{pipi}
\end{figure}

Let us next consider $\pi$--$\pi$ scattering. The diagrams contributing to the scattering amplitudes are depicted in Fig. \ref{pipi}. Due to the presence of a novel contribution coming with a scalar particle in the intermediate state, we expect  a LO correction to the $\chi$PT results. In fact this new contribution allows to test the quark mass dependence of $g_{S\pi\pi}$, the scalar--pion--pion coupling constant, as outlined in \cite{Pelaez:2010fj}. Following this reference, we define $g_{S\pi\pi}^2 := -16 \pi \lim_{s\to m^2_S} (s-m^2_S) t_{00}\,,$ being $t_{00}$ the isospin zero S--wave amplitude. If we take the ratio 
\eqn{gspp}{
\left| \frac{ g_{S\pi\pi}}{ g_{S\pi\pi}^{\rm phys}} \right|={ {\mathaccent 23 m}_{S}^2-(8c_{2m} +2) m_\pi^2  \over (m_{S}^2- 2 m_\pi^2)_{\rm phys} }\,, 
}
and increase the pion mass we find a smooth decreasing function that vanishes around $m_ \pi \approx 368 {\rm MeV}$. While this number roughly matches fit D of Figure 7 of \cite{Pelaez:2010fj}, we do not find the strong quark mass dependence in $g_{S\pi\pi}$ displayed in that figure.

Let us now turn to the evaluation of the scattering lengths.
Their explicit expressions at LO are given by
\begin{eqnarray}
a_0^0=&&  \frac{m^2_{\p}}{\p \Ff^2} \left( \frac{7}{32}-\frac{3}{2} {m^2_{\p}\over 4m^2_{\p}-m^2_{S}} c^2_{1d}+\frac{m^2_{\p}}{ m^2_S }c^2_{1d}\right) \,, \nonumber\\
a_0^2=&&-\frac{m^2_{\p}}{\p \Ff^2}\left( \frac{1}{16}- \frac{m^2_{\p}}{ m^2_{S}}c^2_{1d}\right) \,.  
\label{scln}
\end{eqnarray}
As we have already shown there is a new contribution coming from the scalar exchange. 
Notice that although the size of the $4m^2_\pi-m_S^2$ denominator is similar to $m_S^2$ at the physical point, for certain values of the quark masses this quantity may become small, and hence the self--energy corrections calculated in section \ref{s2p} should be included. However, this will not be necessary for the values of the quark masses considered in this paper. In the sequel we will elucidate the precise role of the scalar in the scattering lengths.

\subsection{First estimates}\label{fes}

Using (\ref{inpp}) and (\ref{cc1d}) into (\ref{scln}) we can compute the values of the scattering lengths. The results, displayed in Table \ref{sl}, show that, for $a_0^0$, $\,\chi$PT$_S$ overshoots the experimental value by roughly the same amount as LO $\chi$PT undershoots it, whereas for $a_0^2$ $\chi$PT$_S$ is roughly a factor of three off the experimental value, namely much worse than LO $\chi$PT, which provides a number pretty close to it already at LO.

This missmatch may be understood as follows. In the decoupling limit (${\mathaccent 23 m}_S^2 \gg m_\pi^2$, $p^2$) the contribution in Fig. \ref{pipi}(b) gives
\be
{\rm Fig.\ref{pipi}\,\, (b)} = - \frac{1}{2} c_{1d}^2 F^2 \langle D_\mu U D^\mu U^\dagger \rangle {1\over -\Box - m^2_{S} } \langle D_\mu U D^\mu U^\dagger\rangle \to
 c_{1d}^2 {F^2\over 2 {\mathaccent 23 m}^2_{S}}  \langle D_\mu U D^\mu U^\dagger \rangle^2\,,
\label{dcl1}
\ee
i.e. it reduces to a contact term which is proportional to $l_1$ in $\chi$PT.  By direct identification one finds the value of the $\chi$PT constant
in terms of the  $\chi$PT$_S$ parameters
\be 
\bar{l}^{\rm LO}_1=192 \p^2\frac{F^2 c^2_{1d}}{{\mathaccent 23 m}_S^2}\,.
\label{l1}
\ee
Note that the usual $4\pi$ suppression factors coming from loop integrals are absent in the tree level calculation above. It is easy to check that the last operator in (\ref{dcl1}) reproduces the scattering lengths (\ref{scln}) in the decoupling limit
\begin{eqnarray}
a_0^0=\frac{m^2_{\p}}{\p \Ff^2} \left( \frac{7}{32}+\frac{5}{2}{m^2_{\p}\over {\mathaccent 23 m}_S^2} c^2_{1d} \right) \,, \quad
a_0^2=-\frac{m^2_{\p}}{\p \Ff^2}\left( \frac{1}{16}-\frac{m^2_{\p}}{{\mathaccent 23 m}_S^2} c^2_{1d} \right) \,.
\label{sclnd}
\end{eqnarray}
Using (\ref{inpp}) leads to $\bar{l}^{\rm LO}_1\sim 38$, roughly $20$ times bigger and with opposite sign than the standard NLO value for this quantity in $\chi$PT, ${\bar l}_1\sim -1.8$ \cite{Colangelo:2001df}. This indicates that a large negative value is expected for $\ell_1$, and, consequently, that NLO contributions are going to be large, at least the ones related to the $\ell_1$ operator. 

\begin{table}
\begin{center}
\begin{tabular}{|c||c|c|} \hline
          & $a^0_0$ & $a^2_0$ \\\hline\hline
Exp.(${\rm stat}$)(${\rm syst}$) & $0.2210(47)(40)$ & $-0.0429(44)(28)$\\ \hline\hline
Beyond NLO $\chi$PT  & $0.220\pm 0.005$ & $-0.0444\pm 0.0010$ \\ \hline\hline
$\chi$PT, LO & $0.159$ & $-0.0454$  \\ \hline
$\chi$PT, NLO & $0.228$ & $-0.0405$  \\ \hline
$\chi$PT$_S$, LO  & $0.275$ & $-0.0121$ \\ \hline
$\chi$PT$_S$, LO+$\ell_1$  & $0.210$ & $-0.0296$ \\ \hline
Linear sigma model & $0.696$ & $-0.0404$ \\ \hline
\end{tabular}
\end{center}
\caption{Values obtained for the scattering lengths from $\chi$PT, $\chi$PT$_S$ and the linear sigma model (bottom panel). 
LO expressions are fixed by the pion mass and decay constant for $\chi$PT, plus the sigma resonance mass for the linear sigma model, plus the sigma resonance width for $\chi$PT$_S$. NLO $\chi$PT values are obtained by fitting ${\bar l}_1+2{\bar l}_2$ and ${\bar l}_3$ to the lattice data of ref. \cite{Fu:2011bz}. LO+$\ell_1$ $\chi$PT$_S$ is obtained by fitting $\ell_1$ and $c_{2m}$ to the same lattice data. In the central panel we show theoretical results that go beyond NLO $\chi$ PT from ref. \cite{Colangelo:2001df}. In the upper panel we show  the values of the scattering lengths extracted from experimental data \cite{Batley:2010zza}.}
\label{sl}
\end{table}

\subsection{Matching with lattice data}\label{matlatd}

The available lattice results for the S--wave scattering lengths use relatively large pion masses, which makes chiral extrapolations less reliable.
In fact, until recently only calculations of $a_0^2$ were available \cite{Yamazaki:2004qb,Chen:2005ab,Beane:2007xs,Feng:2009ij,Dudek:2010ew,Beane:2011sc,Yagi:2011jn}, and the only existing calculation of both $a_0^2$ and $a_0^0$ neglects the disconnected contributions to the latter \cite{Fu:2011bz}. Nevertheless we shall use lattice data of the last reference in order to get a feeling on how $\chi$PT$_S$ performs with respect to the S--wave scattering lengths.

\begin{figure}
  \centerline{
  \includegraphics[width=8cm]{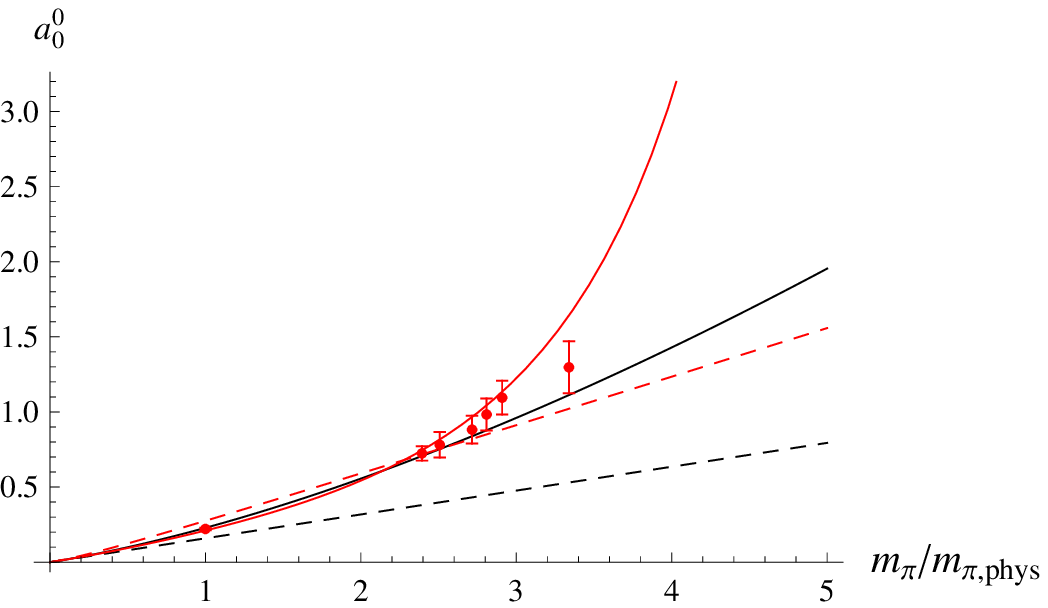}
  \includegraphics[width=8cm]{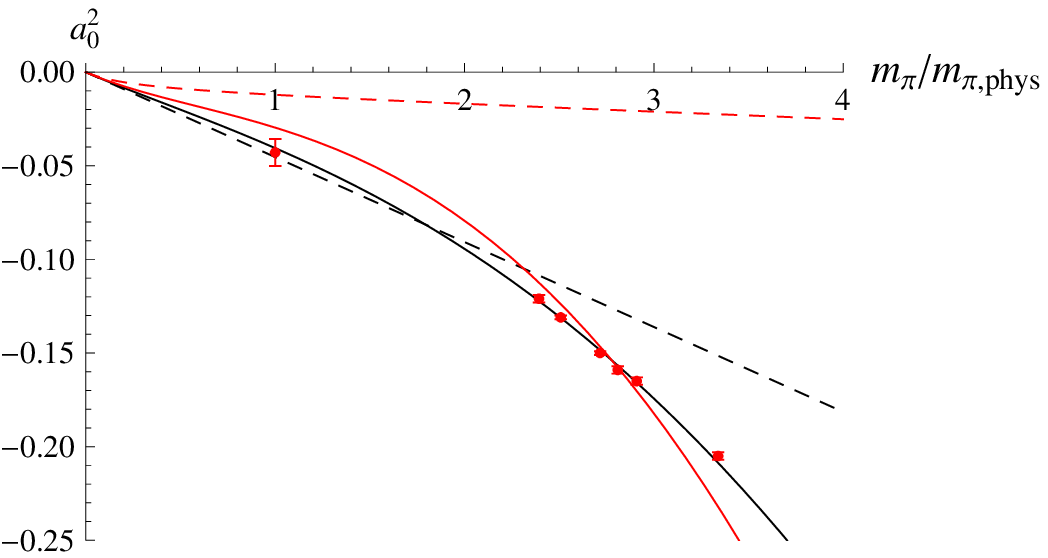}}
  \caption{The best fits of the LO $\chi$PT (black dashed line), NLO $\chi$PT (black solid line), LO $\chi$PT$_S$ (red dashed line) and LO $\chi$PT$_S$ augmented by the operator proportional to $\ell_1$ (red solid line). Red dots are lattice data from \cite{Fu:2011bz}.} 
  \label{lsl}
\end{figure}

As we discussed in section \ref{fes}, the S--wave scattering lengths of $\chi$PT$_S$ at LO are fixed once we input the mass and the width of the sigma resonance in addition to the pion mass and decay constant. Their evolution with the light quark masses is given by that of the pion mass and the LEC $c_{2m}$. By making a combined fit to $a_0^2$ and $a_0^0$ we obtain the dashed red line in Fig. \ref{lsl}. We observed that for $a_0^0$ $\,\chi$PT$_S$ provides a better description of data than LO $\chi$PT (dashed black line), but for $a_0^2$ a much worse one. As argued in section \ref{fes}, large NLO corrections due to ${\ell}_1$ are expected. We may estimated them by just adding its contribution to LO expression. If we fit ${\ell}_1$, we obtain the dashed red line in Fig. \ref{lsl}, and the following numbers
\be
a_0^0=0.210 \quad ,\quad a_0^2=-0.0296 \quad , \quad c_{2m}=-0.443  \quad , \quad {\bar \ell}_1\equiv 96\pi^2 \ell_1 =-16.9 \,.
\ee
Note that we get a large negative number for ${\bar \ell}_1$, consistent with the expectations. Notice also that the value of $c_{2m}$ above justifies the use of formula (\ref{scln}) for $a_0^0$ (i.e. with no self-energy corrections in the $m_S^2-4m_\pi^2$ denominator), with the possible exception of the point corresponding to the largest  pion mass. We see that the description of both scattering lengths improves considerably, the quality of $a_0^0$ being comparable to that of NLO $\chi$PT (black solid line). The plots of NLO $\chi$PT in Fig. \ref{lsl} are obtained by fitting ${ l}_1^r+{ l}_2^r$ and ${ l}_3^r$. The values delivered by the fit are
\be
 {\bar l}_1+2{\bar l}_2=1.4\quad ,\quad {\bar l}_3= -9.3\,,
\ee
which differ quite a lot from the standard values in $\chi$PT at one loop, for instance, ${\bar l}_1+2{\bar l}_2\sim 9.0$ is given in  \cite{Colangelo:2001df} and ${\bar l}_3\sim 2.9$ in \cite{Gasser:1983yg}. In fact if ${\bar l}_3$ is fixed to the last value rather than fitted a very bad description of $a_0^0$ is obtained, whereas the one of $a_0^2$ remains quite good.

The results above encourage us to attempt an extraction of the sigma resonance parameters from the lattice data. We obtain from the fit (to both $a_0^2$ and $a_0^0$)
\be 
c_{2m}=-0.228 \quad ,\quad {\bar \ell}_1=-10.9 \quad ,\quad c_{1d}^2=0.304 \quad ,\quad {\mathaccent 23 m}_S=483 {\rm MeV}\,,
\ee
which produce the following numbers for the sigma decay width and the S--wave scattering lengths
\be
m_S=486 {\rm MeV} \quad ,\quad {\Gamma\over 2}=236 {\rm MeV} \quad ,\quad
a_0^0=0.177 \quad ,\quad
a_0^2=-0.0361 \, .
\ee
The numbers above are quite reasonable for a LO approximation augmented by $\ell_1$, even more if one takes into account that the lattice data is at relatively large pion masses. It shows that our approach may eventually allow for a precise extraction of the sigma resonance parameters from lattice QCD. Note in particular that the value of $c_{2m}$ is compatible with the region of low $\chi^2_{d.o.f.}$ of Fig. \ref{plotscan} and that ${\bar \ell}_1$ remains with a large negative value.

\section{Discussion and Conclusions}\label{discon}

We have considered the possibility that the spectrum of QCD in the chiral limit contains an isosinglet scalar with a mass much lower than the typical hadronic scale $\Lambda_\chi$, and have constructed the corresponding effective theory that includes it together with the standard pseudo--Goldstone bosons, $\chi$PT$_S$. This effective theory has the same degrees of freedom as the linear sigma model, but differs from it in two important points.
First of all, it is conceptually different because the mechanism of spontaneous symmetry breaking is assumed to occur at the scale $\Lambda_\chi$, and hence it is not described within the effective theory. Second, there is a power counting and hence the LO Lagrangian can be augmented at the desired order by adding power suppressed operators. The LO Lagrangian has initially four free parameters more than the linear sigma model, and hence enjoys a larger flexibility to describe data. As explained in the section \ref{chisymcon}, one of these parameters ($c_{1m}$) must be set to zero for consistency, whereas in the linear sigma model it takes a non--zero value. If we force the LO fits to the pion mass and decay constant to go through the linear sigma model values we obtain a $\chi^2_{d.o.f}\sim 135$, namely worse than in LO $\chi$PT$_S$ (which coincides with LO $\chi$PT). Inputing the sigma mass in \cite{Caprini:2005zr}, the linear sigma model delivers a relatively low value for the decay width ($\G /2=188$), a very large value for the isospin zero scattering length ($a_0^0=0.696$) but a pretty reasonable one for the isospin two one ($a_0^2=-0.0404$), see Table \ref{sl}.

At tree level $\chi$PT$_S$ gives definite predictions for S--wave scattering lengths if the mass and decay width of the sigma resonance are used as an input, which are shown in Table \ref{sl}. Neither the value of the isospin zero one ($a_0^0$) nor the one of the isospin two ($a_0^2$) are close to the experimental numbers. Although the value of $a_0^0$ is slightly closer to it than the one obtained in tree--level $\chi$PT, the value of $a_0^2$ is much further away. As argued in section \ref{fes}, this is due to the fact that sizable NLO corrections due to a large value of $\ell_1$ are expected. If we simulate them by letting $\ell_1$ be a free parameter, the combined fits to the lattice data of ref. \cite{Fu:2011bz} to  $a_0^0$ and $a_0^2$ become rather good, see Fig. \ref{lsl}. Note that, although NLO $\chi$PT produces a better description of $a_0^0$ and $a_0^2$ if ${ l}_1^r+2{ l}_2^r$ and ${ l}_3^r$ are fitted to data, the values delivered by the fits of those LECS are incompatible with the ones currently used in $\chi$PT. We have also shown how the combined fits to the S--wave scattering lengths may be used to extract the resonance parameters of the sigma from chiral extrapolations of lattice QCD data.
 
Loop corrections in $\chi$PT$_S$ have been explored in the calculation of $F_\pi$ and $m_\pi$ at NLO. The dynamical scalar field introduces new non--analyticities in the quark mass dependence of these observables, and requires a renormalization of $B$ and $F$, which are absent in $\chi$PT. The fits to the lattice data of ref. \cite{Baron:2009wt} for these observables at NLO in $\chi$PT$_S$ are of similar quality as those at NLO in $\chi$PT. However, when the value of the average light quark masses is extracted from the fit, $\chi$PT$_S$ produces numbers that are closer to those of direct lattice extractions than $\chi$PT does, see Table \ref{qm}. The self--energy of the scalar field has also been calculated at NLO.

We have restricted ourselves to the flavor $SU(2)$ case, the extension to flavor $SU(3)$ is straightforward. In fact because flavor is conserved at any vertex, the contribution to observables with pions involving scalar fields in internal lines are identical and independent of the group, at the order we have calculated. Furthermore, because we will have more parameters at our disposal and $m_S\approx m_K\approx m_\eta$ we expect that the tension between the different contributions to higher chiral orders  \cite{Amoros:1999dp} is alleviated. 

Let us also mention that Lagrangians identical to the first line of (\ref{sigmapion}) are currently being used in the context of composite Higgs models \cite{Contino:2010mh}. In that context, $\chi$PT$_S$ would correspond to an effective theory at the electroweak scale under the assumption that the spontaneous symmetry breaking mechanism takes place at a much higher scale. Small explicit breaking of custodial symmetry at that scale may be taken into account by terms similar to those in the second line of (\ref{sigmapion}).

In summary, we have shown how to consistently introduce a light isosinglet scalar particle in a chiral effective field theory framework, $\chi$PT$_S$. This has consequences concerning the dependence of physical observables on the light quark masses, which have been shown to be compatible with current lattice data. We have also shown that our formalism has the potential to extract the mass and width of the sigma resonance from lattice QCD data. Finally, it would be interesting to explore the consequences of $\chi$PT$_S$ in the chiral approach to nuclear forces \cite{Weinberg:1990rz} (see \cite{Epelbaum:2008ga} for a recent review), since the exchange of a scalar particle is known to be an important ingredient of the nuclear force in one--boson exchange models \cite{Machleidt:2000ge}. 

\bigskip

{\bf Acknowledgments}

\bigskip

We are indebted to Federico Mescia for many discussions, in particular for explanations on the data of ref.\cite{Baron:2009wt}, for bringing to our attention Ref. \cite{Contino:2010mh}, and for the critical reading of the manuscript. We thank Toni Pich for comments on an earlier version
of this work and for bringing to our attention refs. \cite{Rosell:2005ai,Rosell:2006dt,Portoles:2006nr,Rosell:2004mn}. JS also thanks Jos\'e Ram\'on Pel\'aez and Daniel Phillips for comments on earlier versions of this work. We have been supported by the CPAN  CSD2007-00042 Consolider-Ingenio 2010 program (Spain) and the 2009SGR502 CUR grant (Catalonia). JS and JT have also been supported by the FPA2010--16963 project (Spain), and PT by the FPA2010--20807 project (Spain). JT acknowledges a MEC FPU fellowship (Spain).

\section{Appendix}

Through the calculations we have used the following set of  integrals
\begin{equation}
A[m^2_a]=\frac{\Lambda^{4-d}}{i}\int {d^dk\over (2\pi )^d}\frac{1}{k^2-m^2_a+i\epsilon}=m^2_a\left(-2\l+\mu_a\right)\,.
\end{equation}
\begin{equation}
\begin{split}
B[m^2_a,m^2_b;p^2]&=\frac{\Lambda^{4-d}}{i}\int {d^dk\over (2\pi )^d}\frac{1}{k^2-m_a^2+i\epsilon}\frac{1}{(k-p)^2-m_b^2+i\epsilon}=\\
&-2\lambda+{m_a^2\mu_a-m_b^2\mu_b\over \Delta_{ab}}+\bar{J}[m^2_a,m^2_b;p^2]\,.
\end{split}
\end{equation}
Which finite parts are given in terms of  
\eqn{muu}{
\mu_a = - {1\over 16 \pi^2} \ln\left({m^2_a\over \Lambda^2}\right)\,,
}
\eqn{Jbar}{
\bar{J}(m_a^2,m_b^2;p^2)=\frac{1}{32\p^2}\left[2+\left(-\frac{\Delta}{p^2}+\frac{\Sigma}{\Delta}\right)\ln\left(\frac{m^2_a}{m^2_b}\right)-\frac{\n}{p^2}\ln\left(\frac{\left(p^2+\n\right)^2-\Delta^2}{\left(p^2-\n\right)^2-\Delta^2}\right)\right]\,,
}
with $\Delta = \Delta_{ab} = m^2_a-m^2_b$, $\Sigma = m^2_a+m^2_b$, and $\n^2=\left(p^2-\left(m_a-m_b\right)^2\right)\left(p^2-\left(m_a+m_b\right)^2\right)$. The expression in \ref{Jbar} is correct in the momentum region $p^2<(m_a-m_b)^2$. The analytic continuation to higher momentum regions is obtained using the following prescription $p^2\rightarrow p^2+i\epsilon$.
As a last comment, we have used $\overline{\rm MS}$ subtraction scheme
\eqn{}{
\l={1\over 16\pi^2}\left(\frac{1}{d-4}-{1\over 2}[-\g_E+\ln(4\p)+1]\right)\,.
}

\bibliographystyle{ssg}
\bibliography{hair}

\begingroup\raggedright

\endgroup

\end{document}